\documentclass[twocolumn,pra,aps,superscriptaddress,raggedbottom,preprintnumbers,showpacs,floatfix]{revtex4}

\makeatletter\providecommand{\galley@sw}[2]{}\galley@sw{%
\preprint{\fbox{\tiny{\jobname.tex}}}
\usepackage[notref,notcite]{showkeys}
\AtBeginDocument{\addtolength{\hoffset}{0.5\columnwidth}}
}{}\makeatother

\usepackage{amsmath}
\usepackage{amssymb,amsthm}
\usepackage{latexsym}
\usepackage[dvips]{color}
\usepackage{graphicx}
\usepackage{bm}

\renewcommand{\vec}[1]{\mathbf{#1}}
\newcommand{\ii}{\mathrm{i}}
\newcommand{\expup}[1]{\mathrm{e}^{#1}}
\newcommand{\dc}{\delta_{\cal C}}
\newcommand{\thc}{\theta_{\cal C}}
\newcommand{\intc}[1]{\int_{\cal C}\mathrm{d}{#1}\,}

\newcommand{\intlim}[3]{\int_{#1}^{#2}\mathrm{d}{#3}\,}
\newcommand{\nn}{\nonumber}
\newcommand{\mean}[1]{\langle{#1}\rangle}

\newcommand{\ket}[1]{|{#1}\rangle}

\newcommand{\tr}[1]{\mathrm{Tr}{\,#1}}

\renewcommand{\c}{{\cal C}}
\newcommand{\tc}{T_{\cal C}}

\newcommand{\eq}[1]{Eq.~(\ref{#1})}
\newcommand{\fig}[1]{Fig.~\ref{#1}}
\newcommand{\figstwo}[2]{Figs.~\ref{#1} and~\ref{#2}}

\newcommand{\com}[2]{[{#1},{#2}]_{-}}

\newcommand{\eqstwo}[2]{Eqs.~(\ref{#1}) and~(\ref{#2})}
\renewcommand{\tr}[1]{\mathrm{tr}{#1}}
\newcommand{\mixr}{ {\scalebox{1}[1]{\ensuremath{\neg}}} }
\newcommand{\mixl}{ {\scalebox{-1}[1]{\ensuremath{\neg}}} }

\newcommand{\todo}[1]{}

\begin{document}


\title{Auxiliary Hamiltonian representation of the nonequilibrium Dyson equation}

\author{Karsten Balzer}
\affiliation{Max Planck Research Department for Structural Dynamics, University of Hamburg-CFEL,
22607 Hamburg, Germany}
\author{Martin Eckstein}
\affiliation{Max Planck Research Department for Structural Dynamics, University of Hamburg-CFEL,
22607 Hamburg, Germany}


\begin{abstract}
The nonequilibrium Dyson (or Kadanoff-Baym) equation,
which is an equation of motion with long-range memory kernel for real-time Green functions, 
underlies many numerical approaches based on the Keldysh formalism.
In this paper we map the problem of solving the Dyson 
equation  in real-time onto a noninteracting auxiliary Hamiltonian with additional bath degrees 
of freedom. The solution of the auxiliary model does not require the evaluation of a memory kernel 
and can thus be implemented in a very memory efficient way. The mapping is derived for 
a self-energy which is local in space and is thus directly applicable within nonequilibrium dynamical
mean-field theory (DMFT). We apply the method to study the interaction quench in the Hubbard model for an 
optical lattice with a narrow confinement, using inhomogeneous DMFT in combination with second-order 
weak-coupling perturbation theory. We find that, although the quench excites pronounced density 
oscillations, signatures of the two-stage relaxation similar to the homogeneous system can be 
observed by looking at the time-dependent occupations of natural orbitals.
\end{abstract}




\maketitle


\section{Introduction}
The field of strongly correlated materials out of equilibrium is a rapidly 
growing research area. 
On the one hand, this is due to ultrafast pump-probe experiments
which allow one to coherently control and manipulate solids in a time-resolved fashion by external 
laser fields. Examples in this direction include quantum interference effects in photo-excited Mott 
insulators~\cite{wall:11}, light-induced superconductivity in cuprates~\cite{fausti:11}, and 
experiments on ultrafast magnetism~\cite{kirilyuk:10}.
On the other hand, ultracold atomic gases confined in optical lattices~\cite{bloch:08}
allow one to study fundamental condensed matter models for strongly correlated quantum 
systems in great detail, e.g., Ref.~\onlinecite{schneider:12}, and independently of any lattice imperfections. 
In theory, the investigation of correlated systems out of equilibrium has revealed novel 
relaxation phenomena such as doublon decay~\cite{sensarma:10}, 
pre-thermalization~\cite{berges:04,moeckel:08,eckstein:09,stark:13},
and dynamical transitions~\cite{eckstein:09,schiro:10,sciolla:10,tsuji:13}.

The microscopic description of correlated systems out of equilibrium requires appropriate quantum statistical methods. 
A promising approach is provided by nonequilibrium dynamical mean-field theory 
(DMFT)~\cite{freericks:06,review:13}, which works well for higher-dimensional systems and becomes
exact in the limit of infinite dimensions. Other approaches include, e.g., 
cluster perturbation theory~\cite{balzer.m.cpt:11},
linked cluster expansions~\cite{mikelsons:12}, the nonequilibrium dual fermion approach~\cite{jung:12}, 
and nonequilibrium self-energy functional theory~\cite{hofmann:13}. 
All these methods are based on the Keldysh formalism~\cite{keldysh:64} and
involve a Dyson equation which describes the time evolution of a quantum many-body system in 
terms of the one-particle nonequilibrium Green function and a corresponding self-energy~\cite{kadanoff:62}. 
In general, the self-energy introduces time 
retardation effects, which render the numerical solution of the Dyson equation in nonequilibrium a complicated 
task in itself. Therefore, when translational invariance is lost, the solution is restricted to either short times or to 
a small number of orbitals (or bands). Only with a massively parallelized time evolution comprising distributed 
memory~\cite{balzer.prb:10} or with further approximations such as the generalized Kadanoff-Baym 
ansatz \cite{balzer.jpcs:13} these limitations have been overcome so far. 

The idea of the present paper is to develop an alternative method to solve the Dyson equation, which can be efficient and 
computationally less demanding when the self-energy is sufficiently local in space. The approach builds on 
recent work \cite{gramsch:13} 
where it was shown that the action of nonequilibrium DMFT can be mapped onto a single-impurity Anderson 
model by fitting the hybridization function of the DMFT bath. The present paper discusses how 
a similar decomposition of the self-energy defines a noninteracting auxiliary 
Hamiltonian which, on the 
one hand, couples to additional \emph{bath} orbitals but, on the other hand, leads to the same one-particle 
nonequilibrium Green function as the interacting many-body problem we start from. The key to an efficient 
time propagation algorithm lies in the fact
that the auxiliary system involves no interactions, 
such that the corresponding Green function is subject to simple Markovian dynamics and can be determined
by exact diagonalization techniques.
Furthermore, the decomposition of the self-energy is causal (i.e., the time-dependent parameters
of the auxiliary problem depend only on the self-energy at earlier times), such that the mapping can 
easily be incorporated into approaches like nonequilibrium DMFT, where the self-energy 
is given as a functional of the Green function itself.

The paper is organized as follows. In Sections~\ref{subsec:theory-1} and \ref{subsec:theory1}, we describe the theoretical 
framework, elucidate the Dyson equation for the study of nonequilibrium situations, define the auxiliary Hamiltonian and 
formulate the conditions for a valid mapping.
Section \ref{subsec:theory2a} illustrates the mapping within
the Hubbard~I approximation, and Section~\ref{subsec:theory2}  explains in detail the 
decomposition of the self-energy and the determination of the 
parameters in the auxiliary system. Section \ref{subsec:theory4} then
gives details on
the computation of the Green function of the auxiliary model (see also Appendices~\ref{appendixA} and~\ref{appendixB}).
Thereafter, in Section~\ref{sec:scaling}, we test the matrix decomposition of the self-energy for small Hubbard clusters (Sec.~\ref{subsec:results1}), illustrate the 
time-propagation of the auxiliary system (Sec.~\ref{subsec:results1}) and investigate the scalability of the method to long times. 
Finally, Section~\ref{sec:opticallattice} contains our main application. Here, we study the relaxation dynamics of the 
Fermi-Hubbard model following an interaction quench, with a particular focus on the effects of an optical trap.
A summary is presented in Section~\ref{sec:conclusions}.

\section{\label{sec:theory}Theory}

\subsection{\label{subsec:theory-1}Nonequilibrium Dyson equation}

Our main objective is to describe the time-evolution of an interacting quantum many-body system which is initially 
(at time $t=0$) in thermodynamic equilibrium at temperature 
$T=\beta^{-1}$ and evolves unitarily under a 
time-dependent Hamiltonian $H(t)$ for times $t>0$. As prototype we consider the single-band Hubbard model 
\begin{align}
\label{eq:hamiltonian}
 H(t)=&\sum_{ij\sigma}J_{ij}(t)c^\dagger_{i\sigma}c_{j\sigma}+\sum_{i\sigma}(V_{i}(t)-\mu)n_{i\sigma}&\\
 &+U(t)\sum_i(n_{i\uparrow}-\tfrac{1}{2})(n_{i\downarrow}-\tfrac{1}{2})\;,\nn
\end{align}
where $c^\dagger_{i\sigma}$ ($c_{i\sigma}$) are creation (annihilation) operators for an electron with 
spin $\sigma$ on site $i$ of the lattice, $J_{ij}$ denotes the hopping amplitude between sites $i$ and $j$, 
$V_i$ is an external potential, $\mu$ the chemical potential, $n_{i\sigma}=c_{i\sigma}^\dagger c_{i\sigma}$ 
the density and $U$ the local Coulomb interaction.

Using nonequilibrium Green function techniques, 
the time evolution of the Hubbard model (\ref{eq:hamiltonian}) is determined by the Dyson equation
\begin{align}
\label{eq:dyson}
G(t,t')=G_0(t,t')+\!\intc{s}\!\!\intc{\bar{s}}
G_0(t,s)\Sigma(s,\bar s)G(\bar s,t')\;,
\end{align}
where the matrix elements of $G$ are the one-particle nonequilibrium Green functions of 
system (\ref{eq:hamiltonian}) defined on the L-shaped Keldysh time contour $\c$,
\begin{align}
\label{eq:greenfct}
G_{ij\sigma}(t,t')&=-\ii\mean{\tc c_{i\sigma}(t)c^\dagger_{j\sigma}(t')}&\\
&=-\ii\frac{\tr(\tc\{\exp(S)c_{i\sigma}(t)c^\dagger_{j\sigma}(t')\})}{\tr(\tc\{\exp(S)\})}\;,\nn
\end{align}
with action $S=-\ii\intc{t}H(t)$ and contour-ordering operator $\tc$ (see, e.g., Refs.~\onlinecite{kadanoff:62}, \onlinecite{stefanucci:13}, \onlinecite{balzer.lnp:13}, or \onlinecite{review:13}
for an introduction into the Keldysh technique; our notation for contour functions, integrals 
and differentials follows Ref.~\onlinecite{review:13}). Similarly, $G_0$ denotes
the noninteracting Green functions 
$G_{0,ij\sigma}(t,t')=-\ii\mean{\tc c_{i\sigma}(t)c^\dagger_{j\sigma}(t')}_{0}$, evaluated from \eq{eq:greenfct} with $U=0$,
and $\Sigma$ denotes the self-energy with elements $\Sigma_{ij\sigma}(t,t')$. 

The self-energy is typically determined by the Green function in a self-consistent way. Within 
DMFT, for example, $\Sigma(t,t')$ is obtained from the solution of a single-impurity Anderson model 
with a bath that is determined by the lattice Green function. In perturbation theory, the self-energy $\Sigma$ is 
given by a series of Feynman diagrams 
and appears as a functional of 
$G$ and the interaction $U$. Important 
conservation laws such as density, energy and momentum conservation are in particular 
obeyed for any truncation of the derivative $\Sigma_{ij\sigma}(t,t')=\delta\Phi/\delta G_{ji\sigma}(t',t)$, 
where $\Phi[G,U]$ denotes the Luttinger-Ward functional~\cite{baym:62}. Simple examples are 
the Hartree-Fock or second Born approximation which are of first and second order 
in the interaction, respectively.

For a given self-energy, the numerical solution of \eq{eq:dyson} can be performed in different ways. 
One possibility is to discretize all quantities on the time contour $\c$ and to apply standard matrix 
inversion techniques to determine $G$
\cite{freericks:08}. More frequently \eq{eq:dyson} is transformed into a set of integro-differential equations (the Kadanoff-Baym 
equations~\cite{kadanoff:62}) which are then solved within a time propagation scheme, see 
Refs.~\onlinecite{danielewicz:84,koehler:99,stan:09,balzer.prb:10,eckstein:10,tran:08}.
The transformation of \eq{eq:dyson} to differential form is achieved by using the equation 
of motion for $G_0$,
\begin{align}
\sum_{r}[
\delta_{ir}
(\ii\partial_t +\mu)
-
h_{ir}(t)
]
G_{0,rj\sigma}(t,t')
=
\delta_{ij}
\dc(t,t')\;,
\label{eomGG0}
\end{align}
where $h_{ij}$ defines the single-particle part of the Hamiltonian, i.e.,  the quadratic part 
of \eq{eq:hamiltonian} is given by $H_0(t) = \sum_{ij\sigma} (h_{ij}(t)-\mu) c_{i\sigma}^\dagger c_{j\sigma}$. 
In combination with~(\ref{eq:dyson}), \eq{eomGG0} gives
\begin{align}
\sum_{r}
\Big\{
[
\delta_{ir}
&(\ii\partial_t +\mu)
-
h_{ir}(t)
]
G_{rj\sigma}(t,t')
\nonumber\\
&-\intc{s}
\Sigma_{ir\sigma}(t,s)
G_{rj\sigma}(s,t')
\Big\}
=
\delta_{ij}
\dc(t,t')\;.
\label{dysondiff}
\end{align}
This equation clearly reveals  the non-Markovian structure inherent to the 
Dyson equation: The differential $\partial_t G(t,t')$ depends on the value of $G$ at 
different times, and $\Sigma$ takes the role of a memory kernel.
For a self-consistent determination of $\Sigma$ and $G$, the time-propagation of $G$ 
with \eq{dysondiff} and the determination of $\Sigma$ from $G$ can be iterated until 
convergence successively on each time-step. 
A severe restriction for the numerical solution of this equation is the memory needed to 
store the functions $G_{ij\sigma}(t,t')$ for all times on the contour.

\subsection{\label{subsec:theory1}Auxiliary Hamiltonian}

\begin{figure}[t]
\includegraphics[width=0.4825\textwidth]{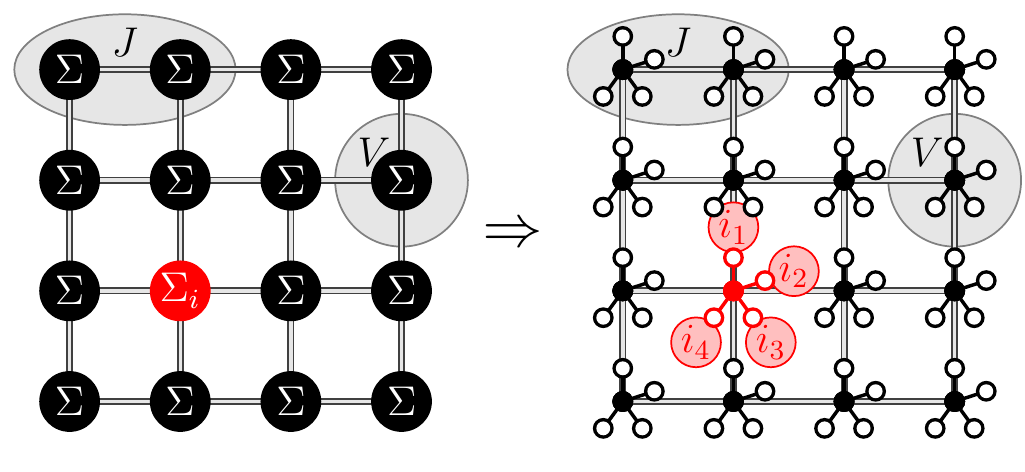}
 \caption{(Color online) Under the assumption of a spatially local self-energy $\Sigma_i$ (compare with \eq{eq:sigma}), the lattice problem (\ref{eq:hamiltonian}) can be mapped onto a noninteracting auxiliary system (right panel) where each lattice site $i$ is coupled to a set of bath orbitals $i_1$, $i_2$, $i_3$ etc. While the large dots in the left hand panel indicate an on-site interaction $U$, the small dots in the right hand panel refer to sites without a local 
 Coulomb interaction; $J$ and $V$ 
 denote the hopping and the external potential which remain the same in the auxiliary system, cf.~\eq{eq:auxhamiltonian}.}
 \label{fig:mapping}
\end{figure}

The central idea of the present paper is to avoid a memory kernel in the time 
propagation scheme for the nonequilibrium Green functions.
To this end, we will map the interacting system (\ref{eq:hamiltonian}) onto a larger auxiliary system  (denoted 
$H_\mathrm{aux}(t)$) which is noninteracting and the Green function of which consequently 
obeys simple Markovian dynamics. The auxiliary system must be constructed such that its 
single-particle Green functions exactly equal the solutions of the Dyson equation \eqref{eq:dyson} 
with a given self-energy. For the derivation below we assume that the self-energy $\Sigma$ is local in 
space,
\begin{align}
 \label{eq:sigma}
 \Sigma_{ij\sigma}(t,t')=\delta_{ij}\Sigma_{i\sigma}(t,t')\;,
\end{align}
which is true for DMFT and thus of wide range of applicability. The generalization of the formalism to non-local self-energies is briefly discussed in the conclusion.

In order to construct the auxiliary Hamiltonian $H_\mathrm{aux}(t)$ we 
connect each individual site $i \equiv i_0$ of the crystal lattice to a set of additional sites 
${\cal B}_i = \{i_1, i_2 ,i_3,...\}$, see \fig{fig:mapping}. We will refer to ${\cal B}_i$ as 
the \emph{bath} (but note that it is different from the bath 
of the effective single-site problem in DMFT).
The additional dynamics between the bath and lattice sites are supposed to 
mimic the retardation effects of the self-energy $\Sigma$. We will see that this is achieved with an auxiliary Hamiltonian that has a quadratic form,
\begin{align}
\label{eq:auxhamiltonian}
 H_\mathrm{aux}(t)=\,&H_0(t)+\sum_{i\sigma}\sum_{l>0}(\epsilon_{i_l\sigma}(t)-\mu)a^\dagger_{i_l\sigma}a_{i_l\sigma}&\\
 &+\sum_{i\sigma}\sum_{l>0}(J_{i_0i_l}^\sigma(t)a^\dagger_{i_l\sigma}c_{i_0\sigma}+\textup{h.c.})\;,\nn
\end{align}
where
\begin{align}
 H_0(t)=&\sum_{ij\sigma}J_{ij}(t)c^\dagger_{i\sigma}c_{j\sigma}+\sum_{i\sigma}(V_{i}(t)-\mu)n_{i\sigma}&\nn\\
 \label{hdef}
\equiv&
\sum_{ij\sigma} (h_{ij}(t)-\mu) c_{i\sigma}^\dagger c_{j\sigma}
\end{align}
is the noninteracting part of \eq{eq:hamiltonian}, the operator 
$a_{i_l\sigma}^\dagger$ ($a_{i_l\sigma}$) creates (annihilates) an electron of spin $\sigma$ on the bath site $i_l$
for $l>0$, $\epsilon_{i_l\sigma}(t)$ are on-site energies of the bath orbitals, and $J_{i_0i_l}^\sigma(t)$ are the additional 
hopping matrix elements between site $i$ and bath orbitals $i_l$ which may depend on the spin.

The time-dependent parameters $J_{i_0i_l}^\sigma(t)$ and $\epsilon_{i_l\sigma}(t)$ must now be chosen
such that the Green functions $G_{i_0 j_0\sigma}^\mathrm{aux}(t,t')$ of the noninteracting model defined by
$H_\mathrm{aux}(t)$ exactly equal the solution of the Dyson equation~\eqref{eq:dyson}, i.e., 
\begin{align}
\label{eq:auxgreenfct}
 G_{ij\sigma}(t,t')=G_{i_0j_0\sigma}^\mathrm{aux}(t,t')\;.
\end{align}
For this purpose, we consider the equations of motion for the Green functions $G_{i_0j_0\sigma}^\mathrm{aux}(t,t')$,
\begin{align}
[\ii\partial_t +\mu]
G_{i_0j_0\sigma}^\mathrm{aux}(t,t')
- 
\sum_{r_0}
h_{i_0r_0}(t)G_{r_0j_0\sigma}^\mathrm{aux}(t,t')
=
\nonumber
\\
\label{eomaux}
\delta_{i_0j_0}
\dc(t,t')
+
\sum_{l>0}
J_{i_0 i_l }^\sigma(t) G_{i_l j_0\sigma}^\mathrm{aux}(t,t')\;,
\end{align}
where $h_{r_0j_0}(t)\equiv h_{rj}(t)$ is defined by \eq{hdef}. Similarly, we can derive an equation of motion for the mixed bath-lattice
term which enters the right hand side of this equation ($l>0$),
\begin{align}
\label{eommix}
[
\ii\partial_t +\mu
-
\epsilon_{i_l\sigma}(t)
]
G_{i_l j_0\sigma}^\mathrm{aux}(t,t')
=
J_{i_l i_0}^\sigma(t) G_{i_0 j_0\sigma}^\mathrm{aux}(t,t')\;.
\end{align}
This equation can be solved by using the Green function 
$g(\epsilon_{i_l\sigma};t,t')$ for an isolated bath orbital with on-site
energy $\epsilon_{i_l\sigma}$, which satisfies
\begin{align}
\label{eomg0}
[
\ii\partial_t +\mu
-
\epsilon_{i_l\sigma}(t)
]
g(\epsilon_{i_l\sigma};t,t')
=
\delta_{\c}(t,t')\;,
\end{align}
and has the explicit form
\begin{align}
\label{eq:bathgreenfct}
g(\epsilon;t,t')
=\ii\left[f_\beta(\epsilon(0)-\mu)-\thc(t,t')\right]
\expup{\ii\!\int_{t}^{t'}\!\!{\mathrm{d}s}[\epsilon(s)-\mu]}
\;.
\end{align}
Here, $f_\beta(\epsilon)=1/(\expup{\beta\epsilon}+1)$ denotes the Fermi-Dirac distribution,
and $\thc$ is the Heavyside step function on the contour. By convoluting \eq{eommix} from the left with 
$g(\epsilon_{i_l\sigma};t,t')$
one obtains
\begin{align}
G_{i_l j_0\sigma}^\mathrm{aux}(t,t')
=
\intc{s}
g(\epsilon_{i_l\sigma};t,s)
J_{i_l i_0}^\sigma(s) G_{i_0 j_0\sigma}^\mathrm{aux}(s,t')\;.
\end{align}
This result can be inserted into Eq.~\eqref{eomaux}, which shows that 
$G_{i_0 j_0\sigma}^\mathrm{aux}(t,t')$ satisfies the equation of motion
\begin{align}
\sum_{r_0}
&
[\delta_{i_0r_0}
(\ii\partial_t +\mu)
-
h_{i_0r_0}(t)
]
G_{r_0j_0\sigma}^\mathrm{aux}(t,t')
\nonumber\\
&-\intc{s}
\Lambda_{i_0\sigma}^\mathrm{aux}(t,s)
G_{i_0j_0\sigma}^\mathrm{aux}(s,t')
=
\delta_{i_0j_0}
\delta_{\c}(t,t')\;,
\label{eomaux1}
\end{align}
with 
\begin{align}
\label{eq:hybridization}
 \Lambda_{i_0\sigma}^\mathrm{aux}(t,t')=\sum_{l>0}J_{i_0i_l}^\sigma(t) 
g(\epsilon_{i_l\sigma};t,t')
 J_{i_li_0}^\sigma(t')\;.
\end{align}
By comparing \eq{eomaux1} with the differential form \eqref{dysondiff} of the Dyson equation 
\eqref{eq:dyson}, we see that the relation \eqref{eq:auxgreenfct} is satisfied, provided 
we can find parameters $J_{i_0i_l}^\sigma(t)$ and $\epsilon_{i_l\sigma}(t)$ such that 
\begin{align}
\label{eq:selfenergymap}
 \Lambda_{i_0\sigma}^\mathrm{aux}(t,t')=\Sigma_{i\sigma}(t,t')\;
\end{align}
for all times $t$ and $t'$ located on the time contour $\c$. We note that condition (\ref{eq:selfenergymap}) must hold only for contributions of 
the self-energy which are beyond the mean-field level while any Hartree contribution can be absorbed 
in an effective potential,
\begin{align}
\label{eq:hartreepotential}
V_{i\sigma}(t)=V_{i}(t)+U(t)(\mean{n_{i\bar\sigma}}-\tfrac{1}{2})\;. 
\end{align}

With \eqstwo{eq:hybridization}{eq:selfenergymap}, the problem of determining the parameters 
of each independent bath ${\cal B}_i$ becomes identical to that of representing a nonequilibrium DMFT 
action by a single-impurity Anderson model, see Ref.~\onlinecite{gramsch:13}. The only difference is that 
instead of the hybridization function of the DMFT bath we here fit the self-energy. In 
Ref.~\onlinecite{gramsch:13}, the existence of solutions and an explicit construction of a solution 
has been discussed.

A short way of summarizing the derivation along the lines of Eqs.~\eqref{eomaux} to \eqref{eomaux1} is to say that the
effective action obtained from the auxiliary model (by integrating out the bath sites) is given by
\begin{align}
\label{eq:auxaction}
 S_\mathrm{aux}=S_0-\ii\sum_{i_0\sigma}\intc{t}\!\!\intc{t'}\Lambda_{i_0\sigma}^\mathrm{aux}(t,t')c^\dagger_{i_0\sigma}(t)c_{i_0\sigma}(t')\;,
\end{align}
where
$S_0=-\ii\intc{t}H_0(t)$
\cite{gramsch:13}. The single-particle Green functions of this quadratic action satisfy the Dyson equation
\eqref{eq:dyson}, provided that \eq{eq:selfenergymap} is satisfied.

\subsection{\label{subsec:theory2a}Application to time-dependent Hubbard I}

In this section we illustrate the approach within the Hubbard I approximation, for 
which the representation~\eqref{eq:selfenergymap} of the self-energy can be derived analytically. 
Within the (non-variational) Hubbard I approximation, the self-energy of the lattice is approximated 
by the self-energy of an isolated Hubbard site with Hamiltonian $H_\text{at}(t)=U(t)n_\uparrow 
n_\downarrow + \sum_\sigma \epsilon_\sigma n_\sigma$. The approximation is the simplest variant 
of the nonequilibrium cluster perturbation theory \cite{balzer.m.cpt:11}, in which the self-energy is 
computed from a small cluster of the lattice. 

For simplicity we consider the case in which the model is driven out of equilibrium only by 
external fields, while the Hubbard interaction is time-independent. The Hubbard I self-energy 
can then be computed from an isolated site in equilibrium. The corresponding Matsubara Green 
function $G^{\text{at}}_\sigma(i\omega_n)$ for the Hamiltonian $H_\text{at}$ is given by
\begin{align}
G^{\text{at}}_\sigma(i\omega_n)
&=
\frac{1-\langle n_{\bar\sigma}\rangle_{\text{at}} }{i\omega_n - \epsilon_\sigma}
+
\frac{\langle n_{\bar\sigma}\rangle_{\text{at}} }{i\omega_n - U-\epsilon_\sigma}\;,
\end{align}
and the self-energy is obtained from inverting
$G^{\text{at}}_\sigma(i\omega_n) = [i\omega_n - \epsilon_\sigma-\Sigma^{\text{at}}_\sigma(i\omega_n)]^{-1}$.
We find
\begin{align}
\Sigma^{\text{at}}_\sigma(i\omega_n)
=
U
\langle n_{\bar\sigma}\rangle_{\text{at}}
+
\frac{a_\sigma^2}{i\omega_n -E_\sigma}\;,
\end{align}
with 
\begin{align}
a_\sigma^2
&= U^2\langle n_{\bar\sigma}\rangle_{\text{at}} \langle 1-n_{\bar\sigma}\rangle_{\text{at}}\;,
\\
E_\sigma
&=
U\langle 1-n_{\bar\sigma}\rangle_{\text{at}}
+\epsilon_\sigma\;.\nn
\end{align}
The analytical continuation of $\Sigma^{\text{at}}_\sigma(i\omega_n)$ to the Keldysh contour
gives
\begin{align}
\label{hubbardIselfenergy}
\Sigma^{\text{at}}_\sigma(t,t')
=
U
\langle n_{\bar\sigma}\rangle_{\text{at}}\dc(t,t')
+
a_\sigma^2
g(E_\sigma;t,t')\;,
\end{align}
where $g(E_\sigma;t,t')$ is given by Eq.~\eqref{eq:bathgreenfct}. 

The time-nonlocal part of the self-energy (\ref{hubbardIselfenergy}) is precisely of the form (\ref{eq:hybridization}).
As a result, solving the Dyson equation with the self-energy $\Sigma^{\text{at}}_\sigma(t,t')$ 
at each lattice site is equivalent to solving the noninteracting lattice problem with {\em only one}
additional bath orbital per lattice site $i$ which is characterized by an on-site energy $\epsilon_{i_1}^\sigma=E_\sigma$
and a time-independent hopping $J_{i_0i_1}^\sigma=a_\sigma$. The 
numerical solution of this single-particle problem involves no memory integrals, and it can 
thus be carried out to arbitrarily large times without any restriction on the memory. A 
similar {\em exact} representation of the self-energy with finitely many bath orbitals is 
possible in general when the (time-dependent) Lehmann representation of $\Sigma$ has 
finitely many terms. This might be useful for certain applications of nonequilibrium cluster 
perturbation theory with small clusters.

\subsection{\label{subsec:theory2}Decomposition of the self-energy}

In general, the representation of the self-energy defined by Eqs.~\eqref{eq:hybridization} and \eqref{eq:selfenergymap}
is not known analytically. To solve \eqstwo{eq:hybridization}{eq:selfenergymap} for the bath parameters, 
we separately consider the various analytical components of the self-energy. In general,  each 
contour function can be parametrized in terms of five components according to different locations of the time arguments on $\c$. 
For the one-particle Green function, we have exemplarily
\begin{subequations}
\begin{align}
 G^<_{ij\sigma}(t,t')&=\ii\mean{c^\dagger_{j\sigma}(t')c_{i\sigma}(t)}\;,&\\
 G^>_{ij\sigma}(t',t)&=-\ii\mean{c_{i\sigma}(t')c^\dagger_{j\sigma}(t)}\;,&\\
 G^\mixr_{ij\sigma}(t,\tau)&=-\ii\mean{c^\dagger_{j\sigma}(\tau)c_{i\sigma}(t)}\;,&\\
 G^\mixl_{ij\sigma}(\tau,t)&=-\ii\mean{c_{i\sigma}(\tau)c^\dagger_{j\sigma}(t)}\;,&\\
 G^\mathrm{M}_{ij\sigma}(\tau)&=-\mean{c_{i\sigma}(\tau)c^\dagger_{j\sigma}(0)}\;,
\end{align}
\end{subequations}
where the argument $t$ ($t'$) is here situated on the upper (lower) real branch of the contour 
and $\tau$ refers to a time on the imaginary track. In addition, we have the Hermitian symmetry 
relations 
\begin{align}
\label{eq:hermsym}
X^\gtrless_{ij\sigma}(t,t')=-[X^\gtrless_{ji\sigma}(t',t)]^*,
\\
X^\mixr_{ij\sigma}(t,\tau)=X^\mixl_{ji\sigma}(\beta-\tau,t)^*,\nn
\end{align} 
for the components of the Green function ($X=G$) and the self-energy 
$(X=\Sigma)$.

While the construction of bath parameters for arbitrary initial states is discussed in detail in 
Ref.~\onlinecite{gramsch:13}, we start in Sections~\ref{sec:scaling} and~\ref{sec:opticallattice} from 
an uncorrelated initial state, i.e., $U(t)=0$ for times $t\leq0$. In this case, the Matsubara and mixed 
components of the self-energy vanish, $\Sigma^\mathrm{M}=\Sigma^\mixl=\Sigma^\mixr=0$, and the remaining 
components of the self-energy are the lesser and greater functions $\Sigma^<$ and $\Sigma^>$ which have 
real time arguments. Following Ref.~\onlinecite{gramsch:13}, we can fit them separately by taking the 
energies of the bath sites entering \eq{eq:bathgreenfct} to be time-independent, i.e., 
$\epsilon_{i_l\sigma}(t)=\mu$ for $t>0$, and by choosing the initial energies $\epsilon_{i_l\sigma}(0)$ 
such that $f(\epsilon_{i_l\sigma}(0)-\mu)$ is either $0$ or $1$. This leads to a representation of the 
self-energy with two sets of bath orbitals, ${\cal B}_i^<$ and ${\cal B}_i^>$, where all sites 
in ${\cal B}_i^<$ (${\cal B}_i^>$) are initially occupied (empty) and ${\cal B}_i={\cal B}_i^<\cup{\cal B}_i^>$. 
More precisely, we have
\begin{align}
\label{eq:dcmplesser}
 -\ii\Sigma^<_{i\sigma}(t,t')=\sum_{l\in{\cal B}_i^<}J_{i_0i_l}^\sigma(t)[J_{i_li_0}^\sigma(t')]^*\;,
\end{align}
and
\begin{align}
\label{eq:dcmpgreater}
 \ii\Sigma^>_{i\sigma}(t,t')=\sum_{l\in{\cal B}_i^>}J_{i_0i_l}^\sigma(t)[J_{i_li_0}^\sigma(t')]^*\;.
\end{align}
In the case of particle-hole symmetry, i.e., for $\mu=0$ in \eq{eq:hamiltonian}, one of the two equations is 
redundant because the greater and lesser functions are then related through 
$\Sigma^<_{i\sigma}(t,t')=\Sigma^>_{i\sigma}(t,t')^*$. If we discretize the times $t$ and $t'$ according to 
$t=t_n=n\Delta t$  and $t'=t_{n'}=n'\Delta t$ with $n,n'\in\{0,1,2,\ldots,N\}$, 
\eqstwo{eq:dcmplesser}{eq:dcmpgreater} have the form of standard matrix decompositions. 
Thus we can obtain an {\em exact} representation of $\Sigma_{i\sigma}$ on the given 
time mesh using in total $L_{i}=2(N+1)$ bath orbitals.

More interesting is the possibility to find an approximate but still accurate representation 
using fewer bath orbitals by applying a suitable low-rank approximation to Eqs.~\eqref{eq:dcmplesser}
and \eqref{eq:dcmpgreater},
\begin{align}
\label{eq:lowrankapprox}
 (-\ii\Sigma^<_{i\sigma})_{nn'}\approx\sum_{l=1}^{L^<_{i}}J_{i_0i_l}^\sigma(t_n)[J_{i_0i_l}^\sigma(t_{n'})]^*\;,
\end{align}
where $L^<_{i}$ is a fixed finite number of bath sites which is smaller 
than the number of time steps $N$ (similarly for $(\ii\Sigma^>_{i\sigma})$).
In the following, we will apply the low-rank Cholesky decomposition to (\ref{eq:dcmplesser}) and 
(\ref{eq:dcmpgreater}) in order to obtain the hopping parameters $J_{i_0i_l}^\sigma(t)$ on the discretized 
time mesh $t=t_n$, which has the advantage of being causal, i.e., the parameters at time $t=m\Delta t$
only depend on the values $(\pm\ii\Sigma^\gtrless_{i\sigma})_{nn'}$ with $n,n'\leq m$.
For technical details concerning the low-rank approximation (\ref{eq:lowrankapprox}) we refer the reader 
to Ref.~\onlinecite{gramsch:13}.

If $L^\gtrless_{i}\ll N$, \eq{eq:lowrankapprox} enables a very compact representation of the self-energy 
where instead of $(N+1)^2$ elements per component $\Sigma_{i\sigma}$ only a small number of $L_{i}(N+1)$ 
elements, namely $L_{i}=L^>_{i}+L^<_{i}$ hopping matrix elements for $N+1$ times, are required to define 
the time dependence of the self-energy. In practice, the numbers $L_{i}^>$ and $L_{i}^<$ act as 
convergence parameters,
and their minimum value depends on the maximum evolution time, cf.~Section~\ref{sec:scaling}. 

\subsection{\label{subsec:theory4}Propagation schemes}

Since the auxiliary model \eqref{eq:auxhamiltonian} is a {\em noninteracting} problem, 
Green functions can be determined by closed equations of motion, cf.~\eq{eomaux}.
In short, we may write
\begin{subequations}
\begin{align}
 \label{eq:kadanoffbaymeqs1}
 \left\{\ii\partial_t+\mu-h_\mathrm{aux}^\sigma(t)\right\} G^\mathrm{aux}_\sigma(t,t')&=
 \dc(t,t')\;,
 \\
 \label{eq:kadanoffbaymeqs2}
  \left\{-\ii\partial_{t'}+\mu-h_\mathrm{aux}^\sigma(t')\right\} G^\mathrm{aux}_\sigma(t,t')&=
 \dc(t,t')\;,
\end{align}
\end{subequations}
where $h_\mathrm{aux}^\sigma(t)$ is the single-particle Hamiltonian of the auxiliary problem, 
and all quantities are viewed as matrices with space and bath orbital indices.
If we label the sites of the crystal lattice with $i=0$, $1$, $2$, $3$ etc.~and let $L=L^<_{i}+L^>_{i}$ 
denote the number of bath orbitals attached to each lattice site (for notational simplicity we assume 
all local self-energies to be represented with the same number of bath orbitals), we can cast the 
single-particle Hamiltonian $h_\mathrm{aux}^\sigma(t)$ into the following 
block matrix form 
(time arguments are omitted),
\begin{align}
\label{eq:haux}
 h_\mathrm{aux}^\sigma(t)=
  \left(
 \begin{array}{cccc}
  a_{00\sigma} & b_{01} & b_{02} & \ldots \\
  b_{10} & a_{11\sigma} & b_{12} & \ldots \\
  b_{20} & b_{21} & a_{22\sigma} & \ddots \\
  \vdots & \vdots & \ddots & \ddots  
 \end{array}
\right)(t)\;,
\end{align}
where all $a$ and $b$ blocks are of dimension $(L+1)\times (L+1)$. While the $a$ blocks in \eq{eq:haux} include the 
hopping to the bath and the effective potential $V_{i\sigma}(t)$, the $b$ blocks involve the hopping terms 
which connect different lattice sites, i.e., (note that by definition $i_0=i$ and $j_0=j$),
\begin{align}
\label{eq:haux1}
  a_{ii\sigma}(t)&=
  \left(
 \begin{array}{cccc}
  V_{i_0\sigma}(t) & J_{i_0i_1}^\sigma(t) & \ldots & J_{i_0i_L}^\sigma(t)\\
  J_{i_1i_0}^\sigma(t) & 0 & \ldots & 0\\
  \vdots & \vdots & \ddots& \vdots\\
 J_{i_Li_0}^\sigma(t) & 0 & \ldots & 0
 \end{array}
 \right)\;,&\\
 \label{eq:haux2}
 b_{ij}(t)&=
 \left(
 \begin{array}{cccc}
  J_{i_0j_0}(t) & 0 & \ldots & 0\\
  0 & 0 & \ldots & 0\\
  \vdots & \vdots & \ddots& \vdots\\
  0 & 0 & \ldots & 0
 \end{array}
 \right)\;.
\end{align}
Note that in the case of nearest-neighbor hopping 
most of the entries in the off-diagonal blocks $b$ vanish, 
and the Hamiltonian $h_\mathrm{aux}^\sigma$ becomes extremely sparse.

If the initial state at time $t=0$ is described by the one-particle density matrix 
$\rho_{i_l j_k\sigma}^\mathrm{aux}(0)=\mean{c_{i_k\sigma}^\dagger(0) c_{i_l\sigma}(0)}_\mathrm{aux}$, the 
solution of \eqstwo{eq:kadanoffbaymeqs1}{eq:kadanoffbaymeqs2} for the
lesser and greater components of the auxiliary Green function gives
\begin{align}
\label{eq:timepropscheme}
 [G^\mathrm{aux}_\sigma]^\gtrless(t,t')=\mp\ii U_\sigma(t,0)R^\gtrless_\sigma U^\dagger_\sigma(t',0)\;,
\end{align}
where $R^>_\sigma=1-\rho_\sigma^\mathrm{aux}(0)$, $R^<_\sigma=\rho_\sigma^\mathrm{aux}(0)$,
and 
\begin{align}
 U_\sigma(t',t)=T_\mathrm{t}\exp\left(-\ii\intlim{t}{t'}{s}h_\mathrm{aux}^\sigma(s)\right)
\end{align}
is the single-particle propagator. 
In Appendix~\ref{appendixA}, the time propagation is explained in detail.

In general, the self-energy is a functional of the Green function (e.g., through the DMFT self-consistency 
and the solution of the impurity problem, or through a self-consistent diagrammatic expansion). 
In \eqstwo{eq:kadanoffbaymeqs1}{eq:kadanoffbaymeqs2}, the self-consistency condition is rather 
hidden in the dependence of the one-particle 
hamiltonian $h^\sigma_\mathrm{aux}=h^\sigma_\mathrm{aux}[G^\mathrm{aux},\Sigma]$ on the auxiliary Green function $G^\mathrm{aux}$ 
(through the Hartree contribution (\eq{eq:hartreepotential}) and the time non-local part of the self-energy $\Sigma$).
In principle, there are two possibilities to obtain self-consistent solutions. On the one hand, we can determine the auxiliary 
Green function for a fixed self-energy for all times and then iterate \eqstwo{eq:kadanoffbaymeqs1}{eq:kadanoffbaymeqs2} 
by  updating the self-energy (and in turn $h^\mathrm{aux}_\sigma(t)$) on the whole time mesh. 
This is easy to implement, but can require
a large number of iterations. On the other hand, we can directly exploit the causality of the Cholesky 
decomposition of the self-energy 
(recall discussion below \eq{eq:lowrankapprox}) 
and set up a time propagation scheme where the self-consistency 
is established on each time slice $n$ separately~\cite{gramsch:13}. In combination with an appropriate (typically 
higher-than-linear order) extrapolation of the hopping matrix elements $J_{i_0i_l}^\sigma(t)$ for times $t\leq t_n$ 
onto the subsequent time slice $n+1$, this guarantees a small number of local iterations which is very advantageous. 
In Appendix~\ref{appendixB}, we describe how one further can apply the Krylov method~\cite{balzer.m:12} to evaluate 
the action of the unitary time evolution operator $U_\sigma(t', t)$ in \eq{eq:timepropscheme} and how the time 
stepping algorithm is in straightforwardly parallelized.

Finally, we mention that the auxiliary bath approach is beneficial also in terms of memory consumption. 
Usually, the numerical solution of the 
Kadanoff-Baym equations is limited by the available computer memory because the total Green function 
$G_{ij\sigma}(t,t')$ is stored in order to evaluate the memory kernel on each time slice~\cite{stan:09}. Long simulations 
with many orbital degrees of freedom require for this reason massive parallelization and a suitable distribution of 
memory over several compute nodes, e.g.,~Refs.~\cite{balzer.prb:10,garny:10}. In the auxiliary bath formalism, on the contrary, it is sufficient to 
store the parameters of the local Hamiltonian \eqref{eq:haux}. 
The number of nonzero parameters is thus determined by the memory needed to store the self-energy $\Sigma_{i\sigma}(t,t')$ which requires by definition considerably less 
memory than the Green function if $\Sigma$ is sufficiently local in space, and if it can 
be represented accurately within a suitable low-rank approximation, cf.~\eq{eq:lowrankapprox}.

\section{\label{sec:scaling}Scaling behavior}

In the following, we test the low-rank decomposition of the self-energy (\eq{eq:lowrankapprox}) and 
illustrate the time propagation of the auxiliary system for two simple cases. 
We will first analyze how well a given self-energy $\Sigma(t,t')$ on a time window $t,t'\leq t_\text{max}$ 
can be represented with a fixed number $L$ of bath orbitals. Subsequently, we will assess how the 
solution of the resulting auxiliary problem converges against the full solution of the Dyson equation 
with increasing $L$.  

To analyze these questions we will use a test self-energy that is generated by 
solving  the Hubbard model (\ref{eq:hamiltonian}) on a small cluster within self-consistent second-order 
perturbation theory, i.e., within the second Born approximation. To be precise, 
the time non-local part of the self-energy is taken to be
\begin{align}
\label{eq:sigma2b}
\Sigma^\gtrless_{i\sigma}(t,t')=U(t)U(t')[G^\gtrless_{i\sigma}(t,t')]^2 G^\lessgtr_{i\sigma}(t',t)\;,
\end{align}
where $G$ are the self-consistent solutions of the Dyson equation \eqref{eq:dyson}. We 
solve Eqs.~\eqref{eq:sigma2b} and \eqref{eq:dyson} either for a single isolated lattice site 
or for a cluster of $2$ by $2$ lattice sites with time-independent nearest neighbor hopping $J_0$. 
To drive the system out of equilibrium, we modify the interaction as a function of time,
\begin{align}
\label{eq:quench}
 U(t)=U_\mathrm{f}\times
 \left\{
 \begin{array}{cc}
 \tfrac{1}{2}(1-\cos(\pi t/t_\mathrm{q}))\;, & t\leq t_\mathrm{q}\\[0.5pc]
 1\;, & t>t_\mathrm{q}
 \end{array}
 \right.\;,
\end{align}
with ramp time $t_\mathrm{q}=2.5$, starting from an uncorrelated state at temperature 
$\beta=10$ and half filling ($\mu=0$). In Figs.~\ref{fig:sigmaerror2x2}~a) and \ref{fig:sigmaerror}~a), we show 
self-consistent reference data for the Green functions and the self-energies for the 
single-site and four-site cluster, respectively.
Below, all energies (times) are measured in units of the (inverse) hopping $J_0$ ($J_0^{-1}$).

Note that we use second-order perturbation theory as an easy way to generate a self-energy 
with the correct analytical properties and a functional form that is similar to self-energies obtained 
within DMFT for large systems: $\Sigma^<(t,t')$ and $\Sigma^>(t,t')$ fall off as a function of 
$t-t'$, but have both nontrivial structure as a function of average time $(t+t')/2$ (due to the 
interaction ramp) and as a function of  $t-t'$, see Figs.~\ref{fig:sigmaerror2x2}~a) and 
\ref{fig:sigmaerror}~a). The true self-energy of a 
single-site cluster is of course different 
and not well described by second-order perturbation theory.

\subsection{\label{subsec:results1}Representation of the self-energy}

\begin{figure}[t]
\includegraphics[width=0.235\textwidth]{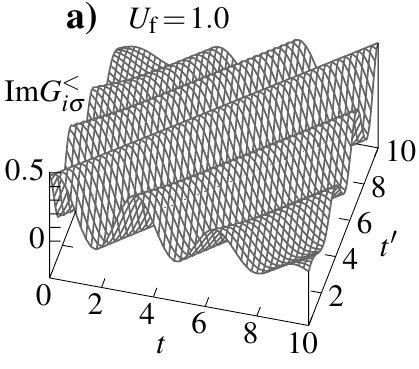}
\includegraphics[width=0.235\textwidth]{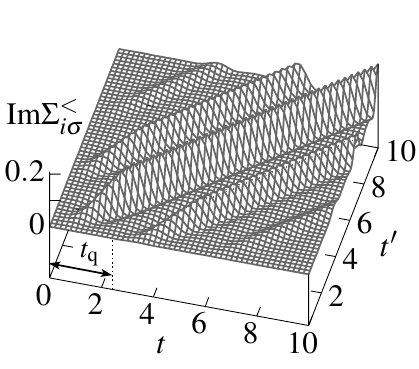}\\
\includegraphics[height=0.2725\textwidth]{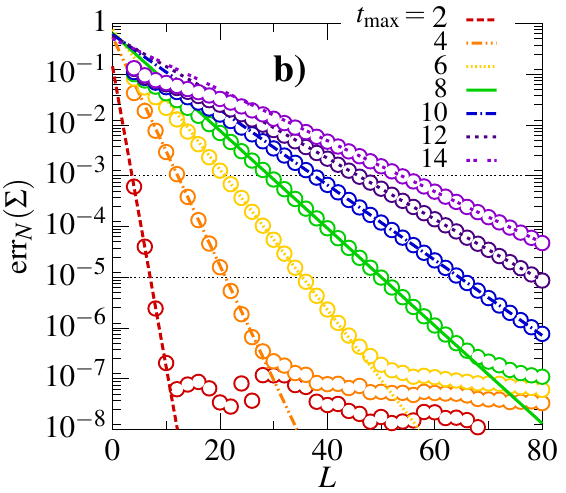}
\includegraphics[height=0.2725\textwidth]{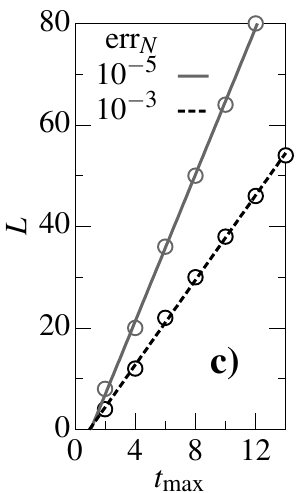}
 \caption{(Color online) a)~Self-consistent Green function $G_{i\sigma}^<$ and local self-energy $\Sigma_{i\sigma}^<$ of the $2$ by $2$ cluster as obtained in second Born approximation from the Dyson equation~\eq{eq:dyson}. The parameters are $\beta=10$, $\mu=0$, $J=J_0$, $t_\mathrm{q}=2.5$ and \mbox{$U_\mathrm{f}=1.0$}. b)~Error of the approximate self-energy evaluated with \eq{eq:lowrankapprox} for various lengths $t_\mathrm{max}$ of the time evolution and different sizes $L$ of the bath. c)~Dependence of the accessible maximum time $t_\mathrm{max}$ on $L$ for a given maximum permitted error $\mathrm{err}_N(\Sigma)$ in the self-energy.}
 \label{fig:sigmaerror2x2}
\end{figure}

To analyze the representation of the self-energy, we compare the given input self-energy 
to the low-rank approximation $\Sigma_\mathrm{approx}$ which is obtained from the matrix 
decomposition (\ref{eq:lowrankapprox}). For the time discretization introduced in 
Section~\ref{subsec:theory2} (i.e., $t_n=n\Delta t$ and $t_{n'}=n'\Delta t$ with $n,n'\in\{0,1,2,\ldots,N\}$), 
we define the corresponding error as
\begin{align}
\label{eq:errordef}
 \mathrm{err}_N(\Sigma)=\sum_{n,n'}\sum_{\alpha\in\{>,<\}}\frac{|\Sigma^\alpha(t_n,t_n')-\Sigma^\alpha_\mathrm{approx}(t_n,t_n')|}{2(N+1)^2}\;.
\end{align}

\fig{fig:sigmaerror2x2}~a) displays the 
input Green function and self-energy for $U_\mathrm{f}=1.0$ for the self-energy of the four-site cluster. 
In \fig{fig:sigmaerror2x2}~b), we plot the error $\mathrm{err}_N(\Sigma)$ of the self-energy decomposition 
\eqref{eq:lowrankapprox} as function of the number of bath orbitals $L$ for different lengths $t_\mathrm{max}$ 
of the time propagation. The size of the time step is thereby fixed to $\Delta t=0.025$, such 
that $N=80$ for $t_\mathrm{max}=2$ and $N=560$ for $t_\mathrm{max}=14$. 
Independent of the value of $t_\mathrm{max}$, we find an exponentially small error for a sufficiently large number of bath sites
[The plateaus for $\mathrm{err}_N(\Sigma)<10^{-6}$ can be attributed to a small number 
$\lambda>0$ (typically $\lambda=10^{-8}$) which we add to 
the diagonal matrix elements $\Sigma(t,t)$ 
of the self-energy in order to  guarantee that the matrices $(-\ii\Sigma^<)$ and $(\ii\Sigma^>)$ 
are positive definite].
On the other hand, we observe that an accordingly larger bath is required in order to reach longer 
simulation times with the same global error $\mathrm{err}_N(\Sigma)$. 

The maximum time which is accessible under a fixed error $\mathrm{err}_N(\Sigma)$
depends linearly on the number of bath orbitals, see \fig{fig:sigmaerror2x2}~c). 
Quantitatively, we find that 
it is sufficient to choose $L$ considerably smaller than the total number of time steps $N$. 
For example, with $40$ bath sites at $t_\text{max}=10$, an error less than $10^{-3}$ is achieved.
On the other hand, $L=40$ bath sites correspond to an effective time step size of $\Delta t=10/40=0.25$ 
which would be too large to obtain numerically converged results in a solution of the integral equation \eqref{eq:dyson}. 
Hence the low-rank decomposition 
has allowed us to effectively compress the information stored in the self-energy $\Sigma(t,t')$.

\begin{figure}[t]
\includegraphics[width=0.235\textwidth]{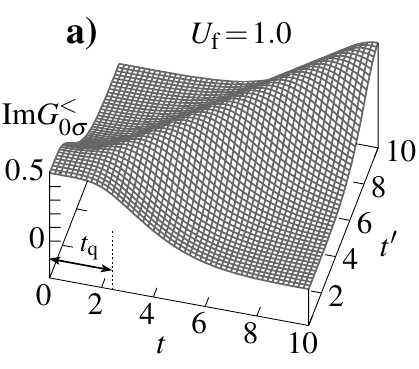}
\includegraphics[width=0.235\textwidth]{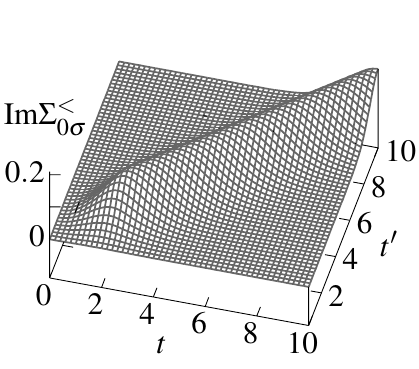}\\
\includegraphics[height=0.2725\textwidth]{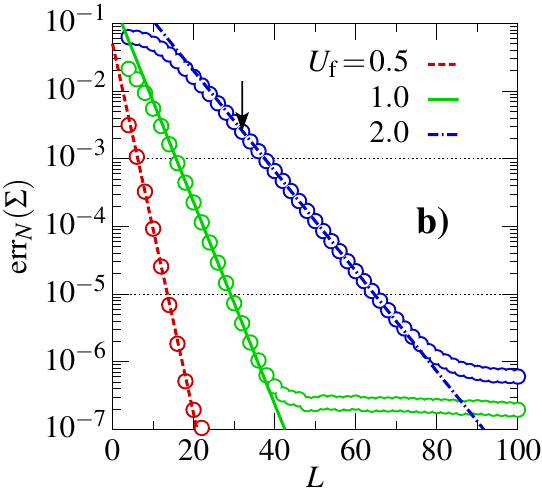}
\includegraphics[height=0.2725\textwidth]{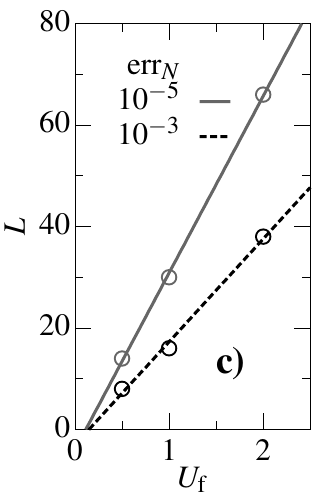}
 \caption{(Color online) a)~Time evolution of the lesser Green function and the lesser self-energy as obtained from \eq{eq:dyson} in second Born approximation for $U_\mathrm{f}=1.0$, $t_\mathrm{q}=2.5$, $\beta=10$ and $\mu=0$. b)~Low-rank Cholesky decomposition of the test self-energy in the interval $[0,t_\mathrm{max}]=[0,10]$ for different parameters $U_\mathrm{f}$. Displayed is the error $\mathrm{err}_{N}(\Sigma)$ as defined in \eq{eq:errordef} with $N=400$ time steps as function of the number of bath orbitals $L$ used in \eq{eq:lowrankapprox}. c)~Scaling of the number of bath sites $L$ with $U_\mathrm{f}$ for fixed errors $\mathrm{err}_N(\Sigma)$.}
 \label{fig:sigmaerror}
\end{figure}

The quality of the representation depends on the functional form of the self-energy. 
We can study this dependence systematically 
for the test self-energy obtained for the isolated site, which has a particularly simple shape: It is characterized by a 
monotonous decay as a function of the difference time $t-t'$ (see Fig.~\ref{fig:sigmaerror}~a)), where the decay time 
decreases with increasing $U_\text{f}$.
\fig{fig:sigmaerror}~b) shows the 
 error $\mathrm{err}_N(\Sigma)$ for different values of $U_\mathrm{f}$ 
 as a function of the total number of bath sites $L$ used in the low-rank Cholesky decomposition. As observed 
 for the four-site self-energy, the error decreases exponentially with increasing number of bath sites,
and $L$ can be chosen smaller than the number of time steps. 
However, the representation of the self-energy for larger $U_\mathrm{f}$ requires 
a larger bath to reach the same level of accuracy. 
\fig{fig:sigmaerror}~c) indicates a linear scaling between the size of the bath and the 
strength of the Coulomb interaction for a given maximum error. 
Taking into account the functional form of $\Sigma$, this 
indicates that the representation of a self-energy which is localized close to the time diagonal needs more bath 
orbitals, which can be understood at least qualitatively: If the self-energy decays to zero for $|t-t'|$ larger than some 
``memory time'' $t_\mathrm{c}$, this can be incorporated into the representation \eqref{eq:lowrankapprox} if each bath site 
is coupled at most for a time period $2t_\mathrm{c}$. Thereafter, new bath sites must be coupled to the system.

\subsection{\label{subsec:results2} Solution of Dyson equation with a low-rank approximation}

\begin{figure}[t]
\includegraphics[width=0.235\textwidth]{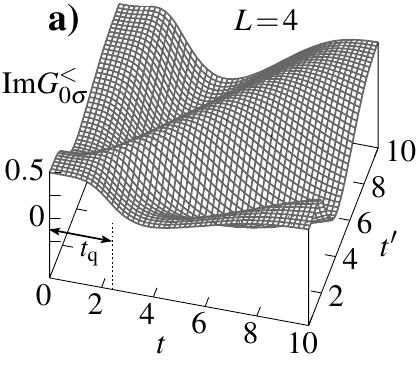}
\includegraphics[width=0.235\textwidth]{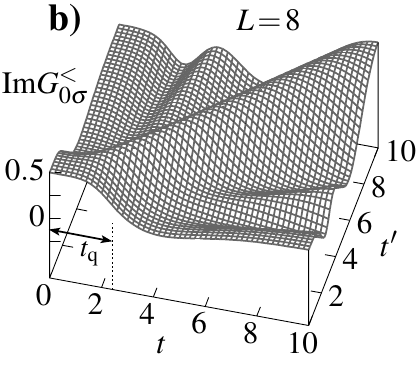}
\includegraphics[width=0.235\textwidth]{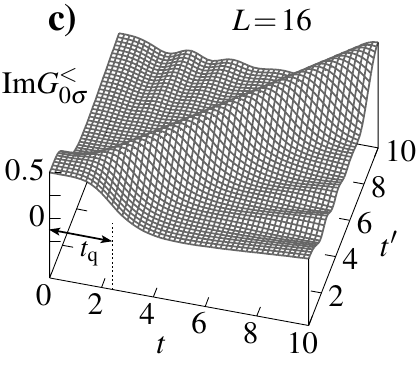}
\includegraphics[width=0.235\textwidth]{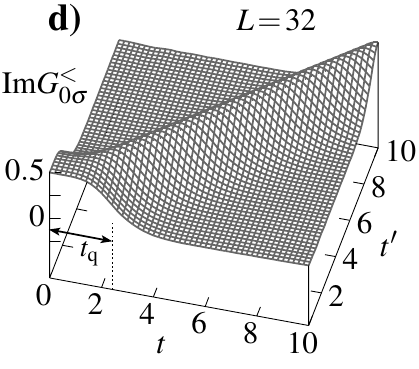}
 \caption{(Color online) Fully self-consistent results for the imaginary part of the Green function $G^{<}_{0\sigma}(t,t')$ for the single-site cluster as obtained in second Born approximation from the time propagation of the auxiliary system for different numbers of bath orbitals $L$. The interaction strength for times $t\geq t_\mathrm{q}$ is $U_\mathrm{f}=2.0$. All other parameters are as in \fig{fig:sigmaerror}, in particular $t_\mathrm{max}=10$ and $N=400$.}
 \label{fig:greenfct}
\end{figure}

As the next step, we demonstrate that the auxiliary bath formalism
is able to reproduce the same Green function $G_{ij\sigma}(t,t')$ as the direct solution 
of \eq{eq:dyson}. To this end, we propagate the auxiliary system (\ref{eq:auxhamiltonian}) in 
time and, following the scheme described in Section~\ref{subsec:theory4}, extract the self-consistent Green 
function for different but fixed sizes of the bath. \fig{fig:greenfct} shows the results for $L=4$, $8$, $16$ and $32$ 
bath orbitals and $U_\mathrm{f}=2.0$ for the single-site cluster (again we use $N=400$ time steps). If the size 
of the bath is too small, we observe that the time evolution of the Green function develops artifacts in form 
of additional oscillations as function of $t$ and $t'$. 
For larger values of $L$, these artifacts shift to later times and finally disappear,
such that the exact solution is well recovered to longer and longer times.
The Green function for $L=32$ is (by eye) barely distinguishable from the exact one.
This is consistent with an error of $\mathrm{err}_N(\Sigma)<10^{-2}$ 
which we find for the self-energy in \fig{fig:sigmaerror}~b), see the black arrow.

\begin{figure}[t]
\includegraphics[width=0.4825\textwidth]{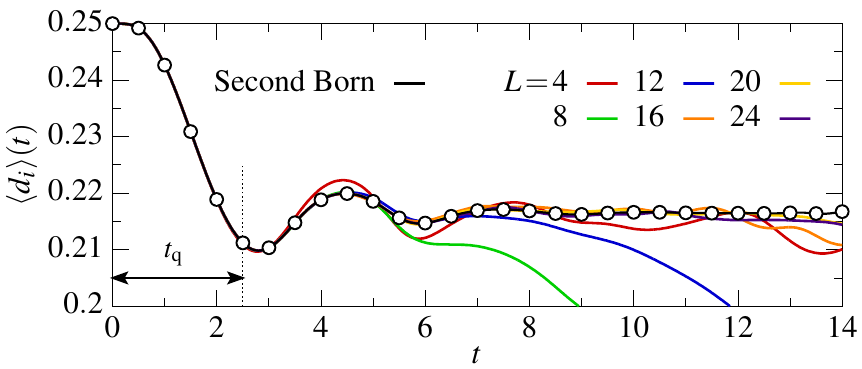}
 \caption{(Color online) Time evolution of the local double occupation $\mean{d_i}(t)$ in the four-site cluster 
 for $U_\mathrm{f}=1.0$ in the second-order Born approximation (black open dots and black solid line). 
The colored curves show the results for different sizes of the bath. 
}
 \label{fig:doubleoccupation2x2}
\end{figure}

The convergence of the low-rank approximation with the number of bath sites can be seen 
even more directly from the time evolution of single-time observables.
In \fig{fig:doubleoccupation2x2}, we exemplarily show results for the local double occupation 
in the four-site cluster for different $L$. The double occupation, which is proportional to the 
interaction energy, is obtained from the convolution
\begin{align}
\mean{d}(t)=&\sum_i\mean{d_i}(t)&\\
=&-\frac{\ii}{U(t)}\sum_i\left\{\intc{s}\Sigma_{i\sigma}(t,s) G_{i_0i_0\sigma}^\mathrm{aux}(s,t')\right\}_{t'=t^+}\;.\nn
\end{align}
The maximum time up to which the 
solution is converged increases with the number of bath orbitals. 
As observed for the representation of the self-energy,
the number $L$ of bath sites required to reach a given accuracy is smaller than the number of time slices 
needed in the conventional solution of the Dyson equation.

\section{\label{sec:opticallattice}2D optical lattice in a harmonic trap}

In this section, we apply the auxiliary Hamiltonian approach 
to investigate the interaction quench in the Hubbard model. 
We particularly focus on the effect of the confinement potential, which is present for experiments with 
ultracold atoms. Interaction quenches in Bose- and Fermi-Hubbard models have been extensively studied in homogeneous systems~\cite{kollath:07,moeckel:08,kollar:08,eckstein:09,
schiro:10}. After a quench from $U=0$ to the weakly interacting regime, 
the system rapidly evolves to a state in which kinetic energy and potential energy are almost 
thermalized, while the momentum distribution function $n(\bm k)$ is still far from its final value. 
In this pre-thermalized state~\cite{berges:04} rapid thermalization is inhibited by 
an infinite number of almost conserved quantities which exist due to the vicinity of the noninteracting 
state~\cite{kollar:11}. Thermalization at longer times and weak coupling is then captured by kinetic 
equations~\cite{moeckel:08,stark:13}.  

Typically, the nonthermal nature of the intermediate state is most clearly evidenced by a discontinuity 
of $n(\bm k)$ across the Fermi surface, which would be absent at any temperature  $T>0$ 
\cite{moeckel:08,eckstein:09,schiro:10}. In the presence of a confinement potential, however,
sharp features like the discontinuity in the momentum occupation are expected to be blurred, and,  
moreover, the interaction quench in a trap might excite collective density oscillations of the atom cloud (e.g., 
a breathing mode), which are superimposed to the relaxation dynamics. The possible observation 
of pre-thermalization in experiment thus requires a good understanding of effects caused by
the trapping potential. Below, we will investigate signatures of a two-stage relaxation for a system 
with a rather narrow confinement, where density oscillations after the quench become very 
pronounced.

\subsection{\label{subsec:optical1}Setup}

\begin{figure}[b]
\includegraphics[width=0.238\textwidth]{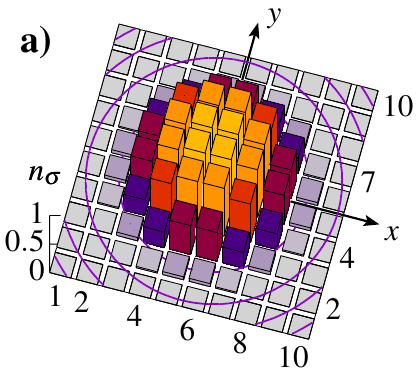}
\includegraphics[width=0.238\textwidth]{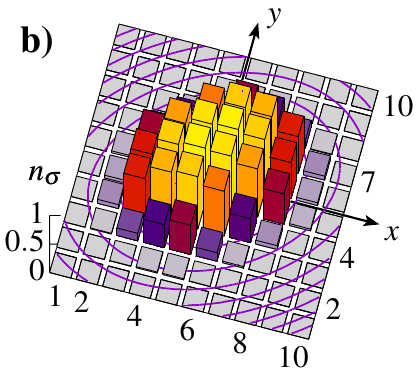}
\includegraphics[width=0.238\textwidth]{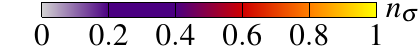}
 \caption{(Color online) Density profiles $n_\sigma(\vec{R}_i)=\mean{n_{i\sigma}}$ for the noninteracting initial states at $\beta=10$ in the rotationally symmetric trap $\mathrm{A}$ (panel a)) and the elongated trap $\mathrm{B}$ (panel b)). In both cases, the total particle number per spin is $\mean{N_\sigma}=20$ ($\mu_\mathrm{A}=3.146$ and $\mu_\mathrm{B}=4.515$). The violet contour lines denote equipotential curves of the time-independent harmonic confinement $V_i$.}
 \label{fig:densityprofile}
\end{figure}

We study the Hubbard model \eqref{eq:hamiltonian} with nearest neighbor hopping
$J_{ij}=\delta_{\langle ij\rangle}J_0$
on a square lattice with $10\times10$ sites.
The optical trap is modeled by a parabolic confinement potential $V_i$ characterized by
two frequencies, $\omega_1$ and $\omega_2$,
\begin{align}
\label{eq:confinement}
V_i(t)=
 \omega_1^2 ({\vec{R}_{i} \hat{\vec{a}}_1})^2+\omega_2^2 ({\vec{R}_{i} \hat{\vec{a}}_2})^2\;.
\end{align}
Here, $\hat{\vec{a}}_j$ are the unit vectors along the principle axes of the trap, and 
the vector $\vec{R}_i$ is pointing from the trap center to the lattice site $i$ 
(the lattice spacing is set to one). 
In the following we compare results for a rotationally symmetric trap with 
$\omega_1^2=\omega_2^2=0.5 J_0$ (referred to as system $\mathrm{A}$, see \fig{fig:densityprofile}~a))
with those for an elongated trap with $\omega_1^2=0.5 J_0$ and $\omega_2^2=J_0$ which is rotated 
by $30$ degrees 
with respect to the lattice (system $\mathrm{B}$, 
\fig{fig:densityprofile}~b)).
The inverse temperature is $\beta=10$, and we fix the average particle number in the trap to 
$\mean{N}=\mean{N_\uparrow}+\mean{N_\downarrow}=40$
by tuning the chemical potential $\mu$ of the initial state. 
The hopping $J_0$ and the inverse hopping $J_0^{-1}$ define the units for energy and time, respectively.
The system is excited by an almost sudden ramp of the electron-electron 
interaction starting from the noninteracting state. The time dependence of the 
quench follows \eq{eq:quench} with $t_\mathrm{q}=0.5$. In all calculations we use 
the DMFT approximation and evaluate the local self-energy in the second-order Born approximation 
(cf.~\eq{eq:sigma2b}).

Before discussing the results, it is interesting to look at the
reduction in computational resource requirements achieved by the auxiliary bath scheme for the 
current problem. For the time grid we choose $N=200$ time steps on the time interval $[0,t_\mathrm{max}]=[0,10]$. 
Within the auxiliary bath scheme, convergence is obtained with $L=64$ bath orbitals at each site of the $10$ 
by $10$ lattice, i.e., the dimension of the associated single-particle Hilbert space is 
$D=10^2(1+64)=6500$. 
An efficient time stepping requires the storage of the auxiliary Hamiltonian in sparse matrix 
form (Eqs.~\eqref{eq:haux1} and \eqref{eq:haux2}) on all time-steps, i.e., approximately 
$6600 \times 200=1,320,000$ 
complex numbers. In contrast, the conventional solution of the Dyson equation would require storing the 
{\em full Green function} for $100$ inequivalent sites and $200$ time steps, which amounts to 
$100^2\times200^2=400,000,000$ complex numbers, taking into account 
all Hermitian symmetries of \eq{eq:hermsym}.

\subsection{\label{subsec:optical2}Time evolution of the density profile and double occupation}

\begin{figure}[t]
\includegraphics[width=0.4825\textwidth]{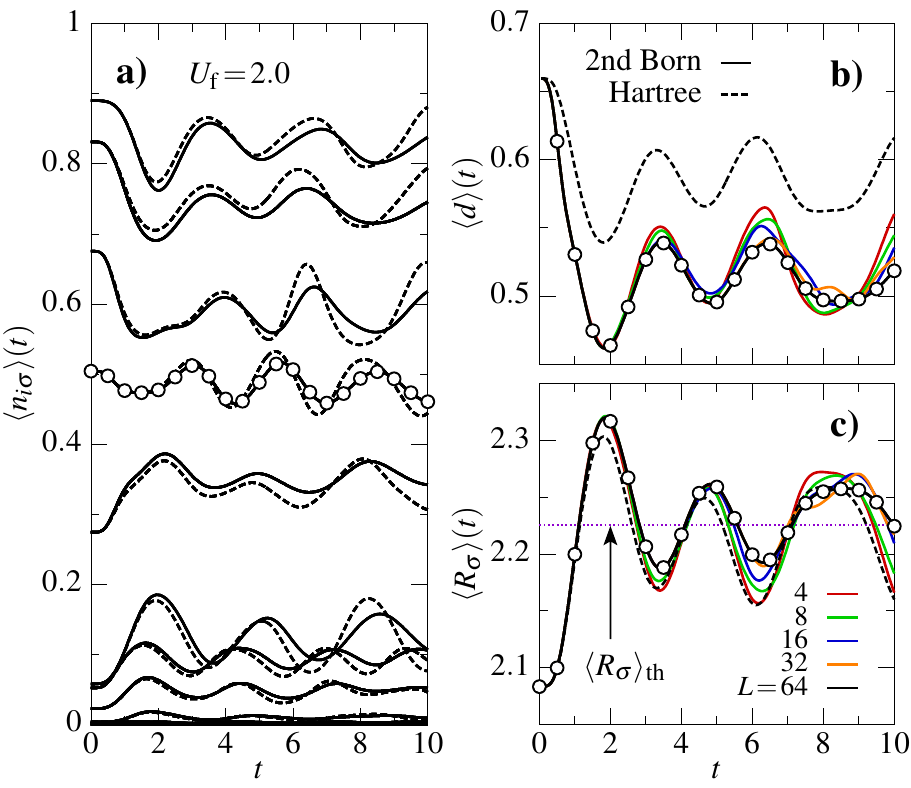}
 \caption{(Color online) Time-dependent observables for the rotationally symmetric trap (system 
 $\mathrm{A}$) and $U_\mathrm{f}=2.0$. 
 a)~site occupations 
 $\mean{n_{i\sigma}}(t)$, b) total double occupation $\mean{d}(t)$ and c) radius $\mean{R_\sigma}(t)$ of the 
 density profile. The black solid (dashed) lines show the result of the second Born (Hartree) approximation. 
 The colored curves indicate the convergence of the results with the size $L$ of the bath in the auxiliary 
 model~(\ref{eq:auxhamiltonian}). In panel c), the violet dotted line refers to the radius $\mean{R_\sigma}_\mathrm{th}$ 
 of an associated thermal equilibrium state (cf.~Section~\ref{subsec:optical2} for discussion). }
 \label{fig:observablesA}
\end{figure}

For times $t\leq0$, i.e., before the switch-on of any interactions, the systems $\mathrm{A}$ and $\mathrm{B}$ are characterized by equilibrium density matrices of the form
\begin{align}
 \rho_{ji\sigma}(0)
 &=
 \langle c_{i\sigma}^\dagger(0)c_{j\sigma}(0)  \rangle 
 \nn\\
\label{eq:initialdensitymatrix}
 &=
 \sum_{\alpha} 
 \langle i\sigma|\alpha\rangle
 \langle\alpha|j \sigma\rangle
f_\beta(\epsilon_{\alpha}-\mu)\;,
\end{align}
where $\epsilon_{\alpha}$  and $\langle i\sigma|\alpha\rangle$ denote the eigenvalues and eigenvectors 
of the associated single-particle Hamiltonian 
(i.e., the matrix elements of \eq{eq:hamiltonian} for $U=0$),
and $f_\beta$ is the Fermi-Dirac distribution.
In \fig{fig:densityprofile}, we show the resulting density profiles 
$n_\sigma(\vec{R}_i)=\mean{n_{i\sigma}}=\rho_{ii\sigma}(0)$,
which are centrally symmetric.
The density of system $\mathrm{A}$ is in addition invariant under rotations of angle $\pi/2$
due to the equal transverse confinements.

For $t>0$, the ramp of the Hubbard interaction $U$ drives the electrons in the traps $\mathrm{A}$ and $\mathrm{B}$ 
out of equilibrium. After the quench, i.e., when $U(t)$ has reached the stationary value $U_\mathrm{f}$, both systems 
evolve unitarily under a new and time-independent Hamiltonian $H'=H(t_\mathrm{q})$. In the course of this, they start 
to redistribute density and double occupation. \figstwo{fig:observablesA}{fig:observablesB} show the time evolution of 
the local densities $\mean{n_{i\sigma}}(t)$ at all sites (see panels a)) as well 
as the total double occupation 
$\mean{d}(t)=\tfrac{1}{\mean{N_\sigma}}\sum_{i}\mean{d_i}(t)$ (see panels b)),
obtained for traps $\mathrm{A}$ and $\mathrm{B}$ at $U_\mathrm{f}=2.0$. 
In addition, in the panels c) we 
plot the time-dependent radius of the density profile $\mean{R_\sigma}(t)$ which is defined by
\begin{align}
 \mean{R_\sigma}^2(t)=\frac{1}{\mean{N_\sigma}}\sum_i\mean{n_{i\sigma}}(t)\vec{R}_i^2\;.
\end{align}

\begin{figure}[t]
\includegraphics[width=0.4825\textwidth]{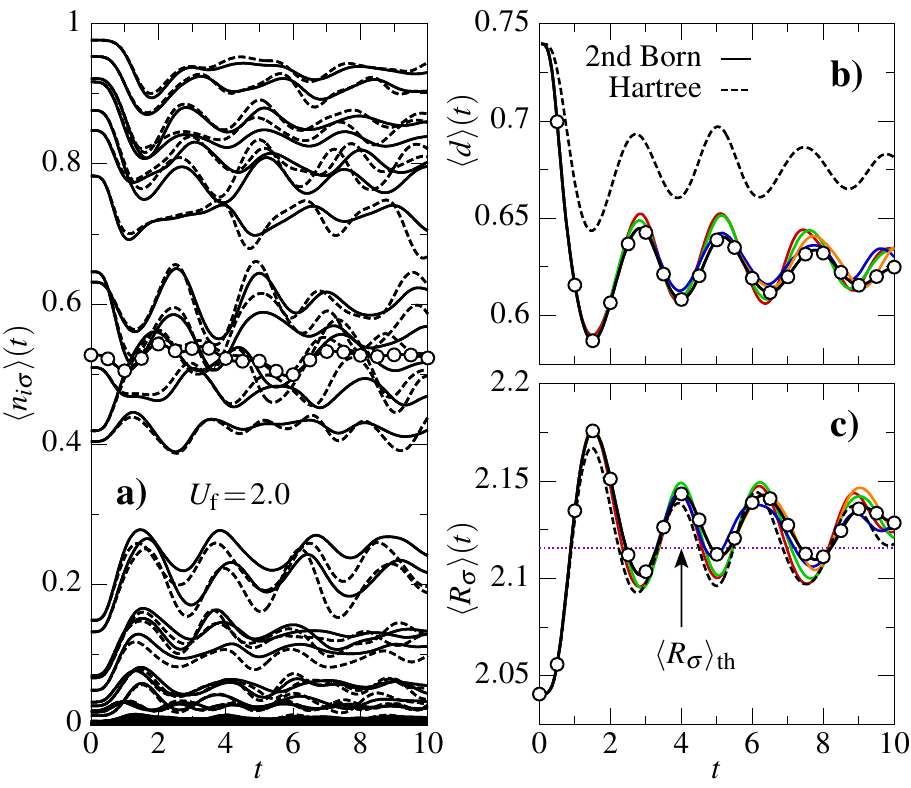}
 \caption{(Color online) Same as \fig{fig:observablesA} but for the elongated trap (system $\mathrm{B}$). The final interaction strength is $U_\mathrm{f}=2.0$.}
 \label{fig:observablesB}
\end{figure}

In Fig.~\ref{fig:observablesA}~a) we can see (for system $\mathrm{A}$) that immediately after the quench the atom cloud spreads out;
Sites of initially high density close to the trap center are depopulated and sites of initially low density 
at the boundary are populated, while densities closer to half-filling exhibit comparatively smaller changes
(open dots). Thereafter, the dynamics 
becomes oscillatory with clearly more than one frequency, which shows that
the system is in a highly excited state after the interaction quench. 
In \fig{fig:observablesB}~a), we identify 
a similar behavior
for the system $\mathrm{B}$. 
In comparison to system $\mathrm{A}$, some of the degeneracies are lifted such that the 
dynamics of the individual densities $\mean{n_{i\sigma}}(t)$ becomes more diverse.
In addition, the increased confinement strength in the direction of $\omega_2$ leads to faster 
oscillations (compare also $\mean{R_\sigma}(t)$ and 
$\mean{d}(t)$ in \figstwo{fig:observablesA}{fig:observablesB}) and to 
a non-uniform redistribution of density. 
The broadening of the density distribution and the subsequent collective oscillation are also well described 
by the time-dependent radius $\mean{R_\sigma}(t)$, see the panels c) in 
\figstwo{fig:observablesA}{fig:observablesB}. Along with the initial expansion of the density, the double occupation decreases in both systems,
cf.~Figs.~\ref{fig:observablesA}~b) and \ref{fig:observablesB}~b). 

In \figstwo{fig:observablesA}{fig:observablesB}  we have also included results obtained within 
the mean-field (Hartree) approximation (black dashed lines). The differences between 
the Hartree and the second Born approximation are more pronounced in $\mean{d}(t)$ than 
in $\mean{n_{i\sigma}}(t)$, while both approximations lead to similar oscillations in the double 
occupation for times $t>2.0$.

In summary, the fast initial change and subsequent oscillations of all observables show that both systems,
$\mathrm{A}$ and $\mathrm{B}$, are not rapidly thermalizing. However, persistent oscillations make it 
hard to identify a pre-thermalization behavior, and it would be useful to find observables that can show 
signatures of a possible two-stage relaxation in a more clear-cut 
way, even for a small and confined system.

\subsection{\label{subsec:optical3}Signatures of pre-thermalization in orbital occupations}

In a homogeneous system, the momentum occupations $n(\bm k)$ provide the 
clearest evidence of pre-thermalization, through 
the discontinuity at the Fermi energy. Yet, for a small system with harmonic 
confinement, the momentum occupations follow a similar diverse and oscillating behavior as the real-space 
densities shown in \figstwo{fig:observablesA}{fig:observablesB}, and a discontinuity in $n(\bm k)$ is absent 
in the spatially inhomogeneous system even at temperature $T=0$. Therefore, a similar analysis of the 
two-stage relaxation as for the homogeneous case 
is rather difficult
for the present systems. On the other hand, regarding the initial state of the system 
at $U=0$, one would still have a discontinuity in the occupations of the single-particle eigenfunctions 
$\ket{\alpha}$ of the trapped system [cf. Eq.~\eqref{eq:initialdensitymatrix}]. This fact motivates to study 
the relaxation in terms of quantities that are more closely related to 
these natural orbitals of the system.

\begin{figure}[t]
\includegraphics[width=0.4825\textwidth]{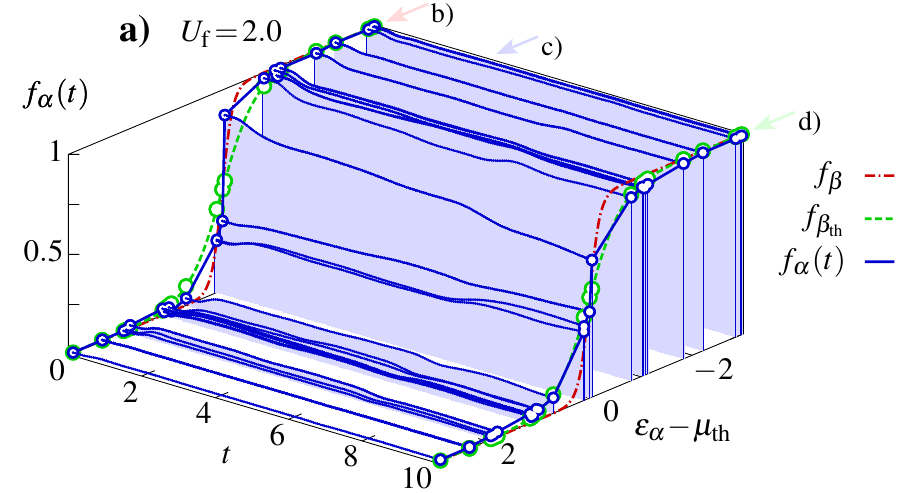}\\[0.5pc]
\includegraphics[width=0.4825\textwidth]{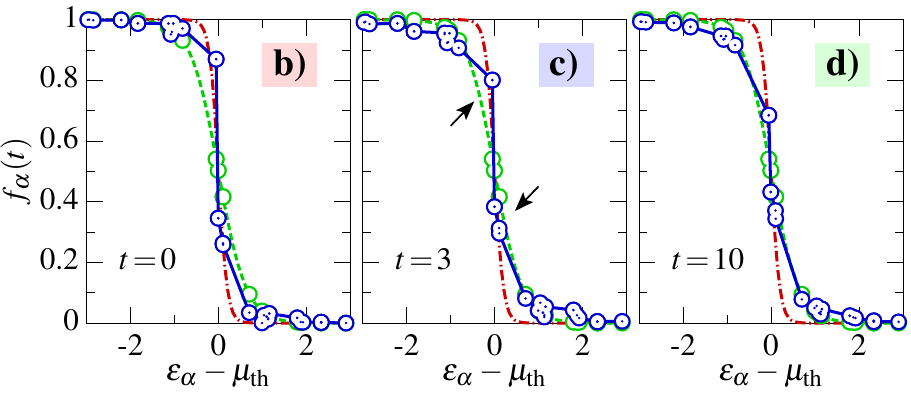}
 \caption{(Color online) Relaxation dynamics in the rotationally symmetric trap $\mathrm{A}$ for a final Coulomb 
 interaction of \mbox{$U_\mathrm{f}=2.0$}. a)~Time 
 evolution of the distribution $f_\alpha(t)$ for the final state basis (blue lines and dots) where $\beta_\mathrm{th}=3.20$ 
 and $\mu_\mathrm{th}=-0.1093$. 
 The panels b), c) and d)~show cuts through the distribution of panel a) at the times $t=0$, $3.0$ and $10$. In all panels, the green dashed line shows the 
 Fermi distribution $f_{\beta_\mathrm{th}}(\epsilon_\alpha^\mathrm{th})$ in the final state basis. As a guide for the eye, we also plot the Fermi
 distribution $f_\beta$ with temperature $\beta=10$ which characterizes the initial state at time $t=0$ (red dash-dotted curve).}
 \label{fig:orbitalrelaxationA}
\end{figure}

From the nonequilibrium Green function $G_{ij\sigma}(t,t')$ of the system, the time-dependent 
distribution function of any given orbital $\ket{\alpha}$ is accessible by
\begin{align}
 f_{\alpha}(t)=-\ii \sum_{ij} 
 \langle\alpha|i\sigma \rangle
G^<_{ij\sigma}(t,t)
\langle j\sigma|\alpha\rangle\;.
\end{align}
In the following, we compare two different natural choices for $|\alpha\rangle$, 
which we refer to as the ``initial state basis'' and the ``final state basis''. The former is 
simply given by the eigenfunctions of the noninteracting (initial) single-particle 
Hamiltonian $h_{ij}=\delta_{\langle ij\rangle}J_0+(V_{i}-\mu)\delta_{ij}$. The 
final state basis will be defined by the eigenbasis of the 
mean-field Hamiltonian 
$(h_\mathrm{th}^\sigma)_{ij}=h_{ij}+U_\mathrm{f}(\mean{n_{i\bar\sigma}}+
\tfrac{1}{2}))\delta_{ij}$, where the effective mean-field temperature $\beta_\mathrm{th}$ 
is computed from the thermal equilibrium Hartree solution which has the same energy and 
particle number as the final state defined by the Green function $G_{ij\sigma}(t,t')$ for $t,t'\geq t_\mathrm{q}$. The corresponding effective temperatures are  
$\beta_\mathrm{th}=3.20$ and $\beta_\mathrm{th}=4.05$ for the quench to $U_\mathrm{f}=2.0$ 
in the systems $\mathrm{A}$ and $\mathrm{B}$, respectively (the adjusted chemical potentials are given in 
\figstwo{fig:orbitalrelaxationA}{fig:orbitalrelaxationB}). Our choice of the single-particle states above is simply motivated by analogy to the 
homogeneous case, where both choices correspond to the plane-wave 
momentum states $|\bm k\rangle$ which well characterize the pre-thermalization 
behavior. 

\begin{figure}[t]
\includegraphics[width=0.4825\textwidth]{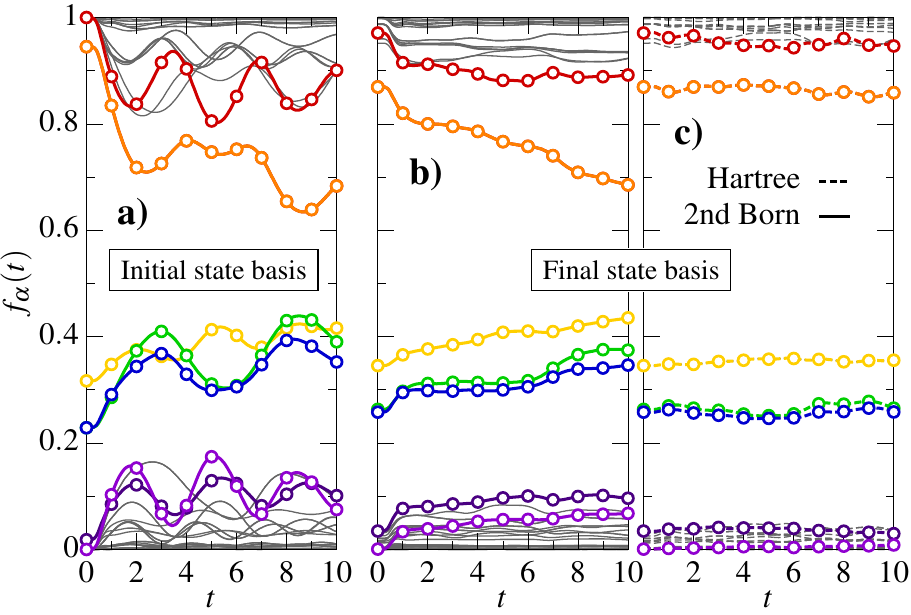}
 \caption{Time-dependent occupations $f_\alpha(t)$ for the ``initial state basis'' (panel~a)) and the ``final state basis'' (panel~b)) for 
 the rotationally symmetric trap $\mathrm{A}$ and $U_\mathrm{f}=2.0$. Panel~c) 
 shows the time evolution of $f_\alpha(t)$ for the final state basis 
 obtained within the  Hartree approximation (see main text).
 }
 \label{fig:orbitalrelaxation01}
\end{figure}

We first analyze the dynamics of the occupations for the rotationally symmetric trap 
$\mathrm{A}$. Figures \ref{fig:orbitalrelaxation01}~a) and b) show
the time-dependent occupations of the initial state basis and the final state basis, respectively. As expected, most occupations correspond
to orbitals $\ket{\alpha}$ that are either fully occupied ($f_\alpha=1$) or almost empty ($f_\alpha=0$) in the initial state. The most pronounced time-dependent 
changes are observed for orbitals close to the Fermi energy (bold colored curves). We find that the occupations of the 
initial state basis still reflect the density oscillations shown in \fig{fig:observablesA}~a) and c). The occupations of the finial state basis, on the other hand, quite clearly reveal 
the two-stage relaxation:
A rapid change of all time-dependent occupations $f_\alpha(t)$ on the time scale of a few inverse hoppings 
(see \fig{fig:orbitalrelaxation01}~b) for times $t\lesssim2.0$) is followed by an almost monotonous drift at
longer times (for the form of the \emph{pre-thermal} distribution as function of the orbital energy, 
see \fig{fig:orbitalrelaxationA}~c)).

It would now be interesting to see whether the drift at long times corresponds to a true thermalization of the 
system. To this end, we in principle would need to compute the (final) interacting equilibrium state with the same amount of 
excitation energy.
For an inhomogeneous system this is quite cumbersome, because multiple calculations are needed to find the 
effective temperature $\beta_\mathrm{th}$ at the correct chemical potential. On the other hand, for small values of 
$U_\mathrm{f}$, a mean-field description is 
usually still quite accurate for equilibrium states, even though higher-order scattering terms are of course crucial to 
correctly describe the actual relaxation dynamics to the thermalized state (this is in line with a description by kinetic 
equations, which  reveals thermalization to a thermal state of the noninteracting system \cite{stark:13}). 
For this reason, it is worthwhile to compare the long-time behavior of the orbital occupations 
$f_\alpha$ to their values in the thermalized mean-field state,
which by construction follow a Fermi distribution 
$f_{\beta_\mathrm{th}}(\epsilon^\mathrm{th}_\alpha)$ at effective temperature $1/\beta_\mathrm{th}$.
Figures \ref{fig:orbitalrelaxationA}~a) to~d) plot the occupations in the final state basis against time and the 
orbital energy $\epsilon^\mathrm{th}_\alpha$. One can see that the drift of the occupations $f_\alpha(t)$
for times $t\gtrsim3.0$ corresponds to a relaxation towards a thermalized state [see in particular the change of the 
occupations close to the Fermi energy from \fig{fig:orbitalrelaxationA}~c) (black arrows) to \fig{fig:orbitalrelaxationA}~d)]. 
This second relaxation process is harder to infer from observables discussed 
in Section~\ref{subsec:optical2},
even taking into account observables that involve averaging over the full trap.
If we compare, e.g., the time evolution of the radius $\mean{R_\sigma}(t)$ 
in the system $\mathrm{A}$ to the thermal value $\mean{R_\sigma}_\mathrm{th}$ 
(see the dotted lines in \fig{fig:observablesA}~c)), 
we observe an oscillation about this value but no clear evidence of damping.

\begin{figure}[t]
\includegraphics[width=0.4825\textwidth]{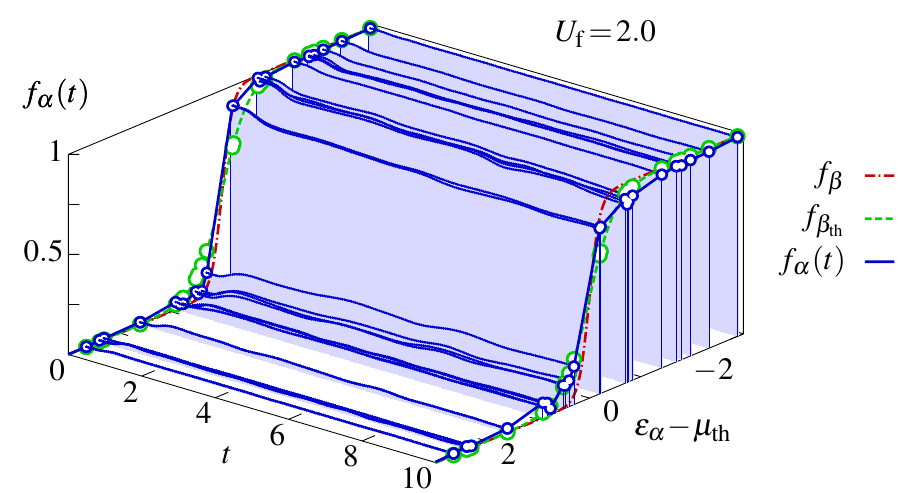}
 \caption{(Color online) Same as in \fig{fig:orbitalrelaxationA}~a) but for the elongated trap $\mathrm{B}$ 
 at $U_\mathrm{f}=2.0$. The effective 
 temperature and the chemical potential in the final state basis are $\beta_\mathrm{th}=4.05$ and 
 $\mu_\mathrm{th}=-0.0026$, respectively.}
 \label{fig:orbitalrelaxationB}
\end{figure}

For the elongated trap (system $\mathrm{B}$), we find a very similar time dependence 
of the distribution function $f_\alpha(t)$, see \fig{fig:orbitalrelaxationB}.
Although there happen to be no single-particle energy levels $\epsilon_\alpha$ very close to the Fermi edge,
we can identify again an intermediate state which the system approaches on a similarly fast time scale,
and further relaxation towards $f_{\beta_\mathrm{th}}$ 
at longer times.

In conclusion, we interpret the presence of the 
intermediate distributions $f_\alpha$ in the 
final state basis around $t=3.0$ as a 
signature that the finite systems $\mathrm{A}$ and $\mathrm{B}$ \emph{pre-thermalize} before 
they actually start to thermalize on a much longer time scale. That this pre-thermalization is mostly driven 
by correlations is demonstrated in \fig{fig:orbitalrelaxation01}~c) where we plot $f_\alpha(t)$ for the 
system $\mathrm{A}$ at $U_\mathrm{f}=2.0$ in Hartree approximation. In contrast to the 
calculation performed in the second-order approximation, the mean-field 
calculation leads to an almost stationary  distribution $f_\alpha(t)$,
even though the 
redistribution of the density $\mean{n_{i\sigma}}(t)$ as discussed in Section~\ref{subsec:optical2} 
is very similar in the Hartree and second Born approximation on the considered time interval
(cf.~\fig{fig:observablesA}~a) and c)). 

\section{\label{sec:conclusions}Summary}

In the present paper, we have formulated a 
method for solving the Dyson equation for an interacting quantum many-body 
system far from equilibrium (\eq{eq:dyson})
which avoids explicit memory integrations 
(or inversions of real-time matrices). Instead, the approach
maps local correlations to an auxiliary bath with finitely many orbitals.
The problem of computing the Green function for the interacting many-body system is thereby 
reduced to an effective single-particle problem or, in other words, to an auxiliary Dyson equation which 
obeys purely Markovian instead of non-Markovian dynamics. 

In Section~\ref{sec:theory}, we have presented the formalism in detail for self-energies which are local in space (single-site DMFT).
The computational benefits of the method are however expected to carry over for a generalization to 
self-energies in cluster DMFT~\cite{maier:05} or cluster perturbation theory~\cite{balzer.m.cpt:11}.
In order to represent a non-local self energy, the additional bath orbitals would be coupled to 
more than one site of the lattice, but the resulting Hamiltonian can still have a simple structure 
provided that the self-energy is sufficiently local in space.
Furthermore, we note that although we have presented only calculations which start from noninteracting thermal states, 
the approach can easily be generalized to correlated initial states. The fundamentals of such an extension are 
formulated in Ref.~\onlinecite{gramsch:13} and lead to the inclusion of further sets of bath orbitals in 
Hamiltonian~(\ref{eq:auxhamiltonian}) which then mimic the decay of initial-state correlations. 

In the context of DMFT, the auxiliary bath approach is most beneficial for
lattice systems which are strongly inhomogeneous in space.
In particular, it has enabled us to 
study an interaction quench for Fermions in an optical lattice,
using inhomogeneous DMFT with second-order perturbation theory as an impurity solver
(without a massive parallelization).
We found that signatures of a two stage relaxation (pre-thermalization followed by slow thermalization)
can be identified in the time-dependent occupations of single-particle orbitals which characterize the 
corresponding thermodynamic equilibrium state, although other observables like the local densities 
exhibits pronounced density oscillations after the quench.
As an obvious extension of this work, one could further substantiating these results with more accurate impurity 
solvers, and study similar questions in the strong coupling regime.

From the computational point of view, 
the efficiency of the auxiliary bath approach partially relies on 
the fact that self-energy decomposition can be more compact than
the conventional representation on an equidistant time mesh.
More precisely, our analysis in Section~\ref{sec:scaling} has shown that 
the number of bath sites can typically 
be chosen smaller than the number of time steps which are propagated. 
Together with the fact that the auxiliary Hamiltonian is anyway very sparse when the self-energy is local,
the compact representation of the self-energy leads over to a tremendous saving of computer memory 
when instead of full Green function only the time-dependent parameters of the auxiliary model are stored.

In further work, it will be interesting to investigate decomposition schemes different from the 
Cholesky decomposition, in order to optimize the representation of the self-energy. In this sense, the auxiliary bath 
provides a starting point to address the issue of systematically truncating memory effects in the Dyson 
equation.

\acknowledgments

We thank Christian Gramsch, Marcus Kollar, Michael Potthoff, and Philipp Werner 
for valuable discussions. Calculations have been performed at the 
PHYSnet computer cluster at University Hamburg.


\appendix

\section{\label{appendixA}Time propagation of the auxiliary Green function}

For the second-quantized quadratic auxiliary Hamiltonian $H_\mathrm{aux}(t)=\sum_{ij\sigma}h_{\mathrm{aux},ij}^\sigma(t) c_{i\sigma}^\dagger c_{j\sigma}$, where $h_{\mathrm{aux},ij}^\sigma(t)$ is given by \eq{eq:haux} and the indices $i,j\in\{1,\ldots,D\}$ run over lattice and bath sites (this is in contrast to the notation in \eq{eq:auxhamiltonian} where we explicitly distinguish between bath and lattice creation and annihilation operators), we need to compute the lesser and greater components of the nonequilibrium Green function
\begin{align}
 G^\mathrm{aux}_{ij\sigma}(t,t')=-\ii\mean{\tc c_{i\sigma}(t)c^\dagger_{j\sigma}(t')}_\mathrm{aux}\;.
\end{align}

To derive an appropriate time-stepping algorithm, we start from the Heisenberg equations of motion for the creation and annihilation operators,
\begin{align}
 \ii\partial_t c_{i\sigma}(t)&=\com{c_{i\sigma}(t)}{H_\mathrm{aux}(t)}\;,&\\
 \ii\partial_t c_{i\sigma}^{\dagger}(t)&=\com{c_{i\sigma}^{\dagger}(t)}{H_\mathrm{aux}(t)}\;,\nn
\end{align}
which in matrix form have the formal solutions,
\begin{align}
 C_\sigma(t)&=U_\sigma(t,0)C_\sigma(0)\;,&\\
C_\sigma^\dagger(t)&=C_\sigma^{\dagger}(0)U^\dagger_\sigma(t,0)\;.\nn
\end{align}
Here, the quantities $C_\sigma$ and $C_\sigma^\dagger$ are row and column vectors of the form $C_\sigma(t)=(c_{1\sigma}(t),\ldots,c_{D\sigma}(t))$ and $C_\sigma^\dagger(t)=(c_{1\sigma}^\dagger(t),\ldots,c_{D\sigma}^\dagger(t))$, and $U_\sigma$ denotes the unitary time evolution operator
\begin{align}
 U_\sigma(t',t)=T_\mathrm{t}\exp\left(-\ii\intlim{t}{t'}{s}h_\mathrm{aux}^\sigma(s)\right)\;,
\end{align}
with the usual time-ordering operator $T_\mathrm{t}$. If the initial state at time $t=0$ is described by the one-particle density matrix $\rho_\sigma^\mathrm{aux}(0)=\mean{C^\dagger_\sigma(0) C_\sigma(0)}_\mathrm{aux}$ which is symmetric, the lesser and greater components of the auxiliary Green function can be computed from
\begin{align}
\label{app:eq:auxgreenfct}
 [G^\mathrm{aux}_\sigma]^\gtrless(t,t')=\mp\ii U_\sigma(t,0)R^\gtrless_\sigma U^\dagger_\sigma(t',0)\;,
\end{align}
where $R^>_\sigma=1-\rho_\sigma^\mathrm{aux}(0)$ and $R^<_\sigma=\rho_\sigma^\mathrm{aux}(0)$ (note that $1$ indicates the identity matrix here). Using the propagator property of $U_\sigma$, we can rewrite \eq{app:eq:auxgreenfct} as
\begin{align}
 [G^\mathrm{aux}_\sigma]^\gtrless(t,t')&=\mp\ii U_\sigma(t,0)R^\gtrless_\sigma U^\dagger_\sigma(s,0)U^\dagger_\sigma(t',s)&\nn\\
&=[G^\mathrm{aux}_\sigma]^\gtrless(t,s)U^\dagger_\sigma(t',s)
\end{align}
or
\begin{align}
 [G^\mathrm{aux}_\sigma]^\gtrless(t,t')&=\mp\ii U_\sigma(t,s)U_\sigma(s,0)R^\gtrless_\sigma U^\dagger_\sigma(t',0)&\nn\\
&=U_\sigma(t,s)[G^\mathrm{aux}_\sigma]^\gtrless(s,t')\;.
\end{align}
Hence, if we choose to propagate the greater correlations function $G^>_\sigma(t_n,t_{n'})$ for times $t_n>t_{n'}$ (we omit the index ``$\mathrm{aux}$'' for simplicity) and the lesser correlation functions $G^<_\sigma(t_n,t_{n'})$ for times $t_n\leq t_{n'}$, where $n,n'\in\{0,1,2,\ldots,N\}$ and $t_0=0$, the algorithm involves the following steps on each time slice $n$ ($m\leq n$) \cite{stan:09}:
\begin{subequations}
\begin{align}
\label{app:eq:steps1}
 G^>_\sigma(t_{n+1},t_{m})&=U_\sigma(t_{n+1},t_n)G^>_\sigma(t_n,t_m)\;,&\\
 \label{app:eq:steps2}
 G^<_\sigma(t_{m},t_{n+1})&=G^<_\sigma(t_m,t_n)U_\sigma^\dagger(t_{n+1},t_n)\;,&\\
 \label{app:eq:steps3}
 G^<_\sigma(t_{n+1},t_{n+1})&=U_\sigma(t_{n+1},t_n)G^<_\sigma(t_n,t_n)U_\sigma^\dagger(t_{n+1},t_n)\;.
\end{align}
\end{subequations}
Note that on the time diagonal it is $G^>_\sigma(t_n,t_n)=-\ii+G_\sigma^<(t_n,t_n)$. To establish the self-consistency directly on the time slice $n$, we further update the time evolution operator $U_\sigma(t_{n+1},t_n)$ a few times by recalculating the single-particle hamiltonian $h^\sigma_\mathrm{aux}$ at the intermediate time $t_n+\tfrac{\Delta t}{2}$. This of course requires a few (low-rank) Cholesky decompositions of the self-energy.

\section{\label{appendixB}Krylov method}

In order to adopt a Krylov-based time propagation scheme~\cite{balzer.m:12} to Eqs.~(\ref{app:eq:steps1}), (\ref{app:eq:steps2}) and (\ref{app:eq:steps3}), we split the matrix multiplications $U_\sigma G^>_\sigma$ and $G^<_\sigma U^\dagger=[U_\sigma (G^<_\sigma)^\dagger]^\dagger$ into sets of matrix-vector multiplications of the form
\begin{align}
\label{app:eq:matrixvectormultiplication}
 U_\sigma G_\sigma=U_\sigma (G_{1\sigma},\ldots,G_{D\sigma})\;,
\end{align}
where $G_{i\sigma}$ denotes the $i$-th column of the matrix $G_\sigma$ which is either $G^>_\sigma$ or $(G^<_\sigma)^\dagger$. For a small time step $\Delta t\ll 1$, each product $U_\sigma(t+\Delta t,t)G_{i\sigma}$ can then be evaluated by applying the Krylov method~\cite{hochbruck:97},
\begin{align}
\label{app:eq:krylov}
 &U_\sigma(t+\Delta t,t)G_{i\sigma}\\
 =&|G_{i\sigma}|\exp\left\{-\ii h^\sigma_\mathrm{aux}(t+\tfrac{\Delta t}{2})\Delta t\right\}\frac{G_{i\sigma}}{|G_{i\sigma}|}&\nn\\
 \approx&|G_{i\sigma}| V_\sigma^{(M)}\exp\left\{-\ii H_\sigma^{(M)}\Delta t\right\}e_1^{(M)}\;,&\nn
\end{align}
where it is essential to first normalize the vectors $G_{i\sigma}$. In the last line of \eq{app:eq:krylov}, 
the matrix $V_\sigma^{(M)}=(V_{1\sigma},\ldots,V_{M\sigma})$ is of dimension $D\times M$ and contains an orthonormal basis of the Krylov space
\begin{align}
{\cal K}^{(M)}=\mathrm{span}(v, h_\sigma v, h^2_\sigma v,\ldots, h^{M-1}_\sigma v)\;,
\end{align}
where $v=G_{\sigma i}$ and $h_\sigma=h_\mathrm{aux}^\sigma(t+\frac{\Delta t}{2})$. Further,
\begin{align}
H^{(M)}_\sigma=[V_\sigma^{(M)}]^* h_\sigma V_\sigma^{(M)} 
\end{align}
is a tridiagonal matrix of dimension $M\times M$ which can easily be diagonalized, and $\mathrm{e}_1^{(M)}$ denotes the first unit vector in $\mathbb{R}^M$. In all practical calculations, a sufficient accuracy is obtained for $M\ll D$.

Finally, we emphasize that the solution of the original lattice problem (\ref{eq:hamiltonian}) requires the computation of the auxiliary Green function $G^\mathrm{aux}_{ij\sigma}(t,t')$ only for indices $i,j$ which are lattice (and not bath) indices, see \eq{eq:auxgreenfct}. This can be exploited to further simplify the time propagation. More precisely, it allows one to evolve $[G^\mathrm{aux}_\sigma]^<$ ($[G^\mathrm{aux}_\sigma]^>$) away from the time diagonal only for those rows (columns) which involve lattice indices, cf.~\eqstwo{app:eq:steps1}{app:eq:steps2}. Along the time-diagonal, such a simplification is inhibited by the specific structure of~\eq{app:eq:steps3} which requires the knowledge of all matrix elements of the Green function.  Furthermore, the time propagation is easily parallelized by performing the independent matrix-vector multiplications in \eq{app:eq:matrixvectormultiplication} simultaneously on many CPUs.





\end{document}